\documentclass[journal=jctcce,
manuscript=article
]{achemso}

\usepackage{graphicx}
\usepackage{dcolumn}
\usepackage{bm}
\usepackage{physics}
\usepackage{soul}
\usepackage{hyperref}
\usepackage{microtype}

\title{
Accuracy and resource advantages of quantum eigenvalue estimation with non-Hermitian transcorrelated electronic Hamiltonians
}

\author{Alexey Uvarov}%
\affiliation{Chemical Physics Theory Group, Department of Chemistry, University of Toronto, Toronto, Ontario M5S 3H6, Canada} \alsoaffiliation{Department of Physical and Environmental Sciences, University of Toronto Scarborough, Toronto, Ontario M1C 1A4, Canada }

\author{Artur F.~Izmaylov}%
\email{artur.izmaylov@utoronto.ca}
\affiliation{Chemical Physics Theory Group, Department of Chemistry, University of Toronto, Toronto, Ontario M5S 3H6, Canada} \alsoaffiliation{Department of Physical and Environmental Sciences, University of Toronto Scarborough, Toronto, Ontario M1C 1A4, Canada }

\begin{document}

\date{\today}

\begin{abstract}
    


    In electronic structure calculations, the transcorrelated method consists in transforming the Hamiltonian so as to remove the Coulomb cusp in its eigenfunctions. As a result, the wavefunction can be described more accurately without increasing the size of the basis set.
    However, the transcorrelated Hamiltonian is non-Hermitian and non-normal, which makes many common quantum algorithms inapplicable. 
    Recently, a quantum eigenvalue estimation algorithm (QEVE) was proposed for non-Hermitian Hamiltonians with real spectra [FOCS~65,~1051~(2024)]. Although the asymptotic scaling of this algorithm with the desired accuracy is shown to be optimal, the constant factor in its complexity scaling has not been analyzed.
    Here we investigate the cost of QEVE applied to transcorrelated electronic Hamiltonians of second-row atoms and compare it to the cost of applying standard qubitization to non-transcorrelated Hamiltonians. 
    We find that, with the xTC approximation, the T gate count of QEVE in the minimal STO-6G basis is between those of standard qubitization in the cc-pVTZ and cc-pVQZ bases. The accuracy of the transcorrelated energy differs between systems: for Li and Be, it is more accurate than the cc-pVQZ energy, while for larger atoms, the error gradually increases, exceeding the cc-pVDZ level for O, F, and Ne.
\end{abstract}

\maketitle

\section{Introduction}

Quantum eigenvalue algorithms for the electronic structure problem can in principle recover the full configuration interaction (CI) energy of the electronic Hamiltonian in a given basis set. The accuracy of the solution is therefore limited by the error of the basis set. For the basis of hydrogenic wavefunctions up to principal quantum number $N$, the error of the basis set scales as $N^{-3}$~\cite{hattigExplicitly2012,helgakerMolecular2014}. In terms of the total number of basis functions $M$, this corresponds to $\sim 1/M$ scaling. The same behavior was observed for Gaussian functions~\cite{klopperinitio1995} and plane waves~\cite{shepherdConvergence2012}. The reason for the slow convergence is that the exact eigenfunctions of the electronic Hamiltonian have cusps at the points where two or more particles coalesce~\cite{katoEigenfunctions1957}. When the two particles are an electron and a nucleus, the cusp is easier to represent accurately: some orbitals, like Gaussian-type orbitals, can be adjusted to account for the cusp~\cite{maScheme2005}, while Slater-type orbitals have a cusp by construction. The electron-electron cusps are more challenging: to accurately reproduce the behavior of the wavefunction at the cusp and in its vicinity, one needs to use large basis sets.

To improve the convergence without resorting to an extremely large CI expansion, conventional computational chemistry has long been taking advantage of explicitly correlated ansatz states---wavefunctions that explicitly depend on the electron-electron distances $r_{ij}$~\cite{hylleraasNeue1929,
kutzelnigg12Dependent1985}. The family of methods relying on this principle, known as F12 theory, is well-developed~\cite{hattigExplicitly2012,gruneisPerspective2017,kongExplicitly2012}. A related approach, called the transcorrelated method, consists in transforming the Hamiltonian with a similarity transformation that depends on $r_{ij}$~\cite{boysdetermination1969,
hirschfelderRemoval1963,ten-nofeasible2000,tsuneyukiTranscorrelated2008}, so that the electron-electron Coulomb repulsion term $r_{ij}^{-1}$ is canceled or at least largely compensated. 
The eigenfunctions of the transformed Hamiltonian can become continuously differentiable (i.e.,~cusp-free); more specifically, the maximum degree of regularity one can obtain this way is $C^{1, 1}$ continuity\cite{fournaisSharp2005}. The similarity transformation can be optimized, e.g., using variational Monte Carlo, so that it not only addresses the cusp condition, but also improves the accuracy of the energy obtained in subsequent calculations~\cite{hauptOptimizing2023}. The Hamiltonian obtained by this procedure, called the transcorrelated (TC) Hamiltonian, can be calculated exactly because the Baker--Campbell--Hausdorff series truncates at the second order. This Hamiltonian is non-Hermitian, and it includes three-body terms. Both of these properties introduce extra challenges for computational chemistry, although the latter is mitigated by the recently introduced xTC approximation~\cite{christlmaierxTC2023,schraivogelTranscorrelated2021,schraivogelTranscorrelated2023}. Another notable way of constructing the transcorrelated Hamiltonian is through the canonical transformation theory of Yanai and Chan~\cite{yanaiCanonical2006}. In this method, the similarity transformation is performed in the second quantized picture, and the generator is an anti-Hermitized projector of the geminal function $f(r_{ij})$. As a result, the canonical transcorrelated Hamiltonian is Hermitian, but it is expressed as an infinite Baker--Campbell--Hausdorff series, which is then approximated further~\cite{yanaiCanonical2012}. 

Recent works have proposed adapting quantum algorithms for the electronic structure problem to leverage the convergence benefits offered by the transcorrelated method. In the quantum computing literature, there are two distinct directions in developing quantum algorithms, each requiring its own approach to TC Hamiltonians. One direction of research focuses on so-called near-term quantum algorithms, aiming to make use of the quantum computers available today or in the near future; such algorithms often rely on some kind of quantum-classical optimization loop and seek to minimize the depth of the quantum circuits used. The other direction in development assumes access to quantum computers with mature error correction, enabling longer circuits. Typically, these algorithms use some form of quantum phase estimation (QPE)~\cite{nielsenQuantumComputationQuantum2010} and thus have a runtime scaling as $O(1/\varepsilon)$ with the desired accuracy $\varepsilon$, as long as one can prepare an input state that has a good overlap with the exact ground state. For comparison, the runtime of variational quantum algorithms typically scales as $O(1/\varepsilon^2)$ per iteration, simply because one needs to evaluate the expected values of the observables by measuring them directly and taking the mean. Thus, despite substantially higher initial requirements, QPE-based algorithms remain potentially more favorable in the long run. Below we briefly review the quantum algorithms for TC Hamiltonians in both near-term and fault-tolerant settings.

Since the CT-F12 framework returns a Hermitian Hamiltonian, existing quantum algorithms for the electronic structure problem like the variational quantum eigensolver~\cite{peruzzoVariationalEigenvalueSolver2014,tillyVariationalQuantumEigensolver2022} can be applied to canonical transcorrelated Hamiltonians without any changes. Motta et al.~performed unitary coupled cluster calculations for small molecules in the CT-F12 framework and reported an improvement in accuracy comparable to increasing the basis quality by one ``zeta''~\cite{mottaQuantum2020}. The $[2]_{\mathrm{R12}}$ perturbative correction of Torheyden and Valeev~\cite{torheydenUniversal2009} can also be used in conjunction with near-term variational algorithms, since we explicitly construct the wavefunction on the hardware and thus have full access to the density matrices~\cite{schleichImproving2022}. Non-Hermitian transcorrelated Hamiltonians require a different treatment. They obey a bivariational principle~\cite{arponenmethod1982,arponenVariational1983,lowdinstability1983}, but not the standard variational principle that is used in variational algorithms. A promising avenue for finding the eigenvalues of non-Hermitian transcorrelated Hamiltonians is through the variational quantum imaginary time evolution (VarQITE) method~\cite{sokolovOrders2023,mcardleImproving2020,dobrautzInitio2024,magnussonEfficient2024}. Finally, one more viable alternative is to use the non-Hermitian variance as a minimization target~\cite{xieVariational2024,solinasBiorthogonal2025}.

In the fault-tolerant regime, the state-of-the-art method for finding eigenvalues relies on implementing a qubitized quantum walk operator and finding its spectrum with quantum phase estimation~\cite{babbushEncoding2018,leeEvenMoreEfficient2021}. While this method would again work without any modification for CT-F12 Hamiltonians, it is not applicable to non-Hermitian TC Hamiltonians: it uses the fact that the left and right eigenvectors of a Hermitian matrix are the same, which is not the case for a non-Hermitian one. However, the problem of finding the eigenvalues of non-Hermitian operators with a quantum computer has been addressed in the literature. For the case when the eigenvalues are real, Shao proposed an algorithm which scales as $O(1/\varepsilon^2)$ with the desired accuracy~\cite{shaoComputing2020}. The recently proposed quantum eigenvalue estimation (QEVE) algorithm~\cite{lowQuantum2024} improves on the former by achieving the optimal $O(1/\varepsilon)$ scaling in the accuracy and linear scaling in the ``one-norm'' of the Hamiltonian presented as a linear combination of unitaries.

Assuming the access to a fault-tolerant quantum computer, the QEVE algorithm could be an adequate method for solving the transcorrelated electronic structure problem due to its asymptotic scaling. However, QEVE is a more complicated algorithm which introduces a certain computational overhead compared to the qubitization algorithm, and without constant-factor analysis it is not clear whether the accuracy improvement associated with the transcorrelated method is going to outweigh the overhead.

In this work, we calculate the resources required to estimate the ground state energy of a transcorrelated Hamiltonian using QEVE and compare them to the costs of standard qubitization. We also estimate the quality of the basis required in standard qubitization to match the accuracy that could be obtained in QEVE. We find that the gate count of QEVE with the TC Hamiltonians is comparable to that of the standard qubitization in the cc-pVQZ basis. The xTC approximation enables a significant reduction in the cost, putting the complexity closer to that of the cc-pVTZ basis. The accuracy of the transcorrelated energy differs between systems: for Li and Be, it is more accurate than the cc-pVQZ energy, while for larger atoms the error gradually increases. Thus, for some systems the QEVE-based workflow showed a significant improvement in the gate count, while for others it showed the same or worse performance. However, the usage of a smaller basis set does enable a reduction in the qubit count for all systems we considered. 

This manuscript is structured as follows. In Section~\ref{sec:theory}, we review the transcorrelated theory and the quantum algorithms for eigenvalue estimation. In section \ref{sec:results}, we present the results on energy and cost estimates for both methods. Section~\ref{sec:conclusions} contains concluding remarks.

\section{Theory}  \label{sec:theory}

\subsection{Background on transcorrelated theory}  \label{sec:tc_theory}

The electronic structure problem consists in finding the eigenvalues and eigenstates of the following Hamiltonian:
\begin{gather}    
    \hat{H}_0 = -\sum_i \frac{\nabla_i^2}{2} + \frac12 \sum_{i\neq j}r_{ij}^{-1} \nonumber \\
    - \sum_{iI}Z_I r_{iI}^{-1} + \frac12 \sum_{I\neq J} Z_I Z_J r_{IJ}^{-1}
    \label{eq:h_original}
\end{gather}
Here lowercase indices count the electrons and uppercase indices count the nuclei. $Z_I$ is the charge of $I$'th nucleus, $r_{ij}$ is the distance between electrons $i$ and $j$, $r_{iI}$ and $r_{IJ}$ are electron-nucleus and nucleus-nucleus distances, respectively. We only consider the electronic wavefunction and treat the coordinates of the nuclei as parameters of the Hamiltonian. We are interested in solving the time-independent Schr\"odinger equation:
\begin{equation}
    \hat{H}_0 \ket{\Psi} = E \ket{\Psi}
\end{equation}

The Hamiltonian~(\ref{eq:h_original}) has singularities at $r_{ij}=0$ and $r_{iI}=0$. Because of that, the eigenfunctions of $\hat{H}_0$ have discontinuous derivatives at these points. More specifically, $\ket{\Psi}$ satisfies the \textit{cusp conditions}~\cite{katoEigenfunctions1957,packCusp1966}. For two electrons with opposite spins, we have
\begin{align}
    \label{eq:cusp}
    \left. \frac{\partial \Psi}{\partial r_{12}}\right|_{r_{12}=0} = \frac12 \left. \Psi\right|_{r_{12}=0}
\end{align}
Here the partial derivative is assumed to be averaged over a sphere centered at the coalescence point. For two electrons with identical spins, we have
\begin{align}
    \left. \frac{\partial \Psi_{1m}}{\partial r_{12}}\right|_{r_{12}=0} = \frac14 \left. \Psi_{1m}\right|_{r_{12}=0} ,
\end{align}
where $\Psi_{1m}$ is the $r_{12} Y_{1m}$ component of $\Psi$ and $Y_{lm}$ is the $(l, m)$-th spherical harmonic. A condition similar to (\ref{eq:cusp}) arises from the electron-nucleus coalescence:

\begin{align}
    \label{eq:cusp_en}
    \left. \frac{\partial \Psi}{\partial r_{iI}}\right|_{r_{iI}=0} = -Z_I \left. \Psi\right|_{r_{iI}=0}
\end{align}

We can represent the eigenfunction $\ket{\Psi}$ as a product of a smooth wavefunction and an operator that reproduces the correct behavior of the wavefunction near the singularities:
\begin{equation}
    \ket{\Psi} = e^{\hat{J}} \ket{\Phi}
\end{equation}
Here $\hat{J}$ is known as the Jastrow factor~\cite{jastrowManyBody1955}. The eigenvalue problem then takes the following form:
\begin{equation}
    \hat{H}_0 e^{\hat{J}} \ket{\Phi} = E e^{\hat{J}} \ket{\Phi}
\end{equation}
Multiplying both sides by $e^{-\hat{J}}$, we obtain an eigenvalue problem for the \textit{transcorrelated} (TC) Hamiltonian~\cite{boysdetermination1969}~$\hat{H}$:
\begin{equation}
\label{eq:tc_hamiltonian}
    \hat{H} \ket{\Phi} = e^{-\hat{J}} \hat{H}_0 e^{\hat{J}} \ket{\Phi} = E \ket{\Phi}
\end{equation}
The Jastrow factor generally has the following form:
\begin{equation}
    \hat{J} = \sum_{i<j} \tau(\mathbf{r}_i, \mathbf{r}_j) \equiv \sum_{i<j} \tau_{ij},
\end{equation}
where $\mathbf{r}_i$ are the electron coordinates, and $\tau(\mathbf{r}_i, \mathbf{r}_j) = \tau_{ij}$ is symmetric with respect to its arguments. In principle, $\tau_{ij}$ depends not only on the electron-electron distance $r_{ij}$, but also on the electron-nucleus distances $r_{iI}$.

The similarity transformation, expanded as Baker--Campbell--Hausdorff series, truncates exactly after two commutator terms:
\begin{equation}
    e^{-\hat{J}} \hat{H}_0 e^{\hat{J}} = \hat{H}_0 - [\hat{J}, \hat{H}_0] + \frac12 [\hat{J}, [\hat{J}, \hat{H}_0]]
\end{equation}
because of the nilpotency of the Heisenberg algebra generated by the coordinate and momentum operators. That is, using the Leibniz rule, one can reduce the expressions $[\hat{J}, \nabla_i^2]$ to expressions involving only $[x_i, \frac{\partial}{\partial x_i}] = -1$. Concretely, here $\hat{H}_0$ is a function of the coordinates and their second derivatives, the commutator $[\hat{J}, \hat{H}_0]$ only depends on the coordinates and the first derivatives, and $[\hat{J}, [\hat{J}, \hat{H}_0]]$ does not depend on the derivatives at all, and therefore commutes with $\hat{J}$.
By calculating the commutators, we obtain $\hat{H}$:
\begin{gather}
    \hat{H} = \hat{H}_0 - \sum_i \left(
    \frac{\nabla_i^2 \hat{J}}{2} + (\nabla_i \hat{J}, \nabla_i) + \frac12 (\nabla_i \hat{J})^2
    \right) \nonumber \\
    = \hat{H}_0 - \sum_{i< j}\hat{K}(\mathbf{r}_i, \mathbf{r}_j) - \sum_{i<j < k} \hat{G}(\mathbf{r}_i, \mathbf{r}_j, \mathbf{r}_k)
\end{gather}
where 
\begin{align}
    \hat{K}(\mathbf{r}_i, \mathbf{r}_j) & = \frac12 \left(\nabla_i^2 \tau_{ij} + \nabla_j^2 \tau_{ij} + (\nabla_i \tau_{ij})^2 +  (\nabla_j \tau_{ij})^2\right) \nonumber \\
    &+ (\nabla_i \tau_{ij}, \nabla_i) + (\nabla_j \tau_{ij}, \nabla_j) \\
    \hat{G}(\mathbf{r}_i, \mathbf{r}_j, \mathbf{r}_k) &= (\nabla_i \tau_{ij}, \nabla_i \tau_{ik}) \nonumber \\
    &+ (\nabla_j \tau_{ji}, \nabla_j \tau_{jk}) + (\nabla_k \tau_{ki}, \nabla_k \tau_{kj})
\end{align}

After this transformation, we project the Hamiltonian onto a finite basis of Slater determinants to obtain a second-quantized Hamiltonian:

\begin{gather}
    H = \sum_{pq, \sigma} h_{pq} a^\dagger_{p \sigma} a_{q \sigma} \nonumber \\
    + \sum_{pqrs, \sigma \tau} (V_{pqrs} - K_{pqrs}) a^\dagger_{p \sigma} a^\dagger_{q \tau} a_{s \tau} a_{r \sigma} \nonumber \\
    +
    \sum_{pqrsuv, \sigma \tau \kappa} G_{pqrsuv} a^\dagger_{p \sigma} a^\dagger_{q \tau} a^\dagger_{r \kappa} a_{v \kappa} a_{u \tau} a_{s \sigma}
\label{eq:tc_h_sq}
\end{gather}
Here $h_{pq} = \bra{\phi_p}\frac{-\nabla^2}{2} - \sum Z_Ir_{1I}^{-1} \ket{\phi_q}$, $V_{pqrs} = \bra{\phi_p \phi_q}r_{12}^{-1}\ket{\phi_r \phi_s}$, $K_{pqrs} = \bra{\phi_p \phi_q}\hat{K}\ket{\phi_r \phi_s}$, $G_{pqrsuv} = \bra{\phi_p \phi_q \phi_r}\hat{G}\ket{\phi_s \phi_u \phi_v}$. The Hamiltonian now has three-body interactions, but more importantly, the two-body term is non-Hermitian.
In addition, even though the full space TC Hamiltonian~(\ref{eq:tc_hamiltonian}) is diagonalizable, the projected Hamiltonian~(\ref{eq:tc_h_sq}) is not guaranteed to retain this property, nor is it guaranteed to have real eigenvalues.  As an example of how a projected non-Hermitian operator could lose diagonalizablity, the following matrix:
$$
\begin{pmatrix}
    0 & 1 & 2 \\
    0 & 0 & 3 \\
    4 & 5 & 6
\end{pmatrix}
$$
is diagonalizable, but its upper-left $2 \times 2$ block is not. Nonetheless, up to numerical precision, all projected Hamiltonians we considered in this work had real spectra and were diagonalizable.

The number of terms in the Hamiltonian~(\ref{eq:tc_h_sq}) scales as $O(M^6)$ with the number of molecular orbitals $M$. To slow down this scaling, one can construct an effective two-body Hamiltonian using the xTC approximation~\cite{christlmaierxTC2023,schraivogelTranscorrelated2021,schraivogelTranscorrelated2023}. The approximation is implemented by putting the Hamiltonian in the generalized normal order~\cite{kutzelniggNormal1997} and discarding all three-body terms that remain in this representation. 
The numerical results of Refs.~\cite{schraivogelTranscorrelated2021,schraivogelTranscorrelated2023} suggest that even in the simplest form of the xTC approximation, where the generalized normal order is defined with respect to the Hartree--Fock state, the error is negligible compared to the usual error sources like basis set incompleteness or the limitations of a particular computational method. 

\subsection{Ground state energy estimation for Hermitian and non-Hermitian operators} \label{sec:qeve_theory}

In this section we briefly review the main ideas behind the estimation of the lowest eigenvalue of a Hamiltonian using a quantum computer. The algorithms for both Hermitian and non-Hermitian Hamiltonians assume that we represent the Hamiltonian as a linear combination of unitaries:
\begin{equation}
    H = b_0 + \sum_{j=1}^K b_j U_j
\end{equation}
Here $b_j$ are real and positive, with the phases being absorbed in the unitaries. The important complexity metrics are the number of terms $K$ and the one-norm
\begin{equation}
    \alpha := \sum_{j=1}^K b_j
\end{equation}
Here the constant term $b_0$ is not included in the count as we can perform all quantum operations without it and then add $b_0$ to the final results.

The Hamiltonian $H$ is represented in a quantum computer using the PREPARE and SELECT circuits which act on the target register and a $\lceil \log K \rceil$-qubit ancilla register as follows:
\begin{subequations}
\begin{gather}
    \mathrm{PREP} |0\rangle \ket{\psi} = \frac{1}{\sqrt{\alpha}} \sum \sqrt{b_j}\ket{j} \ket{\psi} \\
    \mathrm{SEL} \ket{i}\ket{\psi} = \ket{i} U_i \ket{\psi}
\end{gather}
\end{subequations}
Here and in the rest of the paper, $\log K$ denotes a logarithm with base two.
Applying $\mathrm{PREP}^\dagger \cdot \mathrm{SEL} \cdot \mathrm{PREP}$ to an input state $\ket{0}\ket{\psi}$ then yields a state
\begin{equation}
    \ket{0} \frac{H}{\alpha} \ket{\psi} + \ket{\perp},
\end{equation}
where $\ket{\perp}$ is orthogonal to any state with zero in the ancilla register.

\subsubsection{Hermitian operators}

We first consider the case when $H$ is a Hermitian operator with eigenstates $\{\ket{\lambda_i}\}$ and corresponding eigenvalues $\{\lambda_i\}$. The core component of the algorithm for ground state energy estimation is a controlled unitary $U$ that encodes the eigenvalues of $H$ in its eigenphases.
For a Hermitian $H$, there are two common choices of such a unitary:
\begin{itemize}
    \item \emph{Trotter-based simulation:} construct an approximate propagator
    \[
        U(t) \approx e^{-iHt},
    \]
    so that $U(t) \ket{\lambda_j} = e^{-i \lambda_j t} \ket{\lambda_j}$~\cite{whitfieldSimulationElectronicStructure2011}.
    \item \emph{Qubitization:} construct a walk operator $W$ such that the states $\mathrm{PREP} |0\rangle \ket{\lambda_i}$ and $\mathrm{SEL}\cdot \mathrm{PREP} |0\rangle \ket{\lambda_i}$ form a two-dimensional invariant space for $W$, and the eigenvalues of $W$ restricted to this subspace are $\exp(\pm i \arccos{\lambda_i/\alpha})$~\cite{lowHamiltonian2019,leeEvenMoreEfficient2021}.
\end{itemize}
The ground state energy of a Hermitian operator $H$ is then estimated as follows. Here we show the steps for the Trotter-based algorithm, but for qubitization they are the same up to substitution of the input state by $\mathrm{PREP} \ket{0} \ket{\psi_0}$ and the eigenstates of $H$ by the eigenstates of $W$.

\textit{Input state preparation.} We start by preparing an input state $\ket{\psi_0}$ which approximates the ground state $\ket{\lambda_0}$. In the eigenbasis, $\ket{\psi_0}$ has the following decomposition:
\begin{equation*}
        \ket{\psi_0} = \sum_{j} c_j \ket{\lambda_j}
\end{equation*}
The success probability of the algorithm is proportional to the overlap $|c_0|^2 = |\braket{\psi}{\lambda_0}|^2$, so it is desirable to maximize it. 

\textit{Controlled applications of the phase unitary.} Introduce an $m$-qubit ancilla register in the uniform superposition
    \[
        \frac{1}{2^{m/2}} \sum_{k=0}^{2^m -1} \ket{k}.
    \]
    Apply controlled-$U^k$, i.e., apply $U^k$ to the system register when the ancilla is in state $\ket{k}$. The joint state becomes
    \[
        \frac{1}{2^{m/2}} \sum_{k=0}^{2^m-1} \ket{k} \, U^k \ket{\psi_0}.
    \]
    Expansion of $\ket{\psi_0}$ in the eigenbasis yields
    \[
        \frac{1}{2^{m/2}} \sum_{k=0}^{2^m-1} \sum_{j} c_j \ket{k} \, e^{-i k \lambda_j t} \ket{\lambda_j}.
    \]
\textit{Quantum Fourier transform on the ancilla register.} Finally, we apply the quantum Fourier transform to the ancilla register:
    \[
        \ket{k} \mapsto \frac{1}{2^{m/2}} \sum_{y=0}^{2^m-1} e^{2\pi i k y / 2^m} \ket{y}.
    \]
    After QFT, the ancilla register is approximately peaked around $y \approx \theta_j 2^m / (2\pi)$. Measuring the ancilla gives an estimate of $\theta_j$, and thus of $\lambda_j$, with probability $|c_j|^2$. In particular, with probability $|c_0|^2$ the outcome encodes the ground state energy.

\subsubsection{Non-Hermitian operators}

When $H$ is non-Hermitian, one cannot build the phase-encoding unitary the same way as for the Hermitian case. The evolution operator $e^{-iHt}$ is no longer unitary, and the walk operator does not have the same two-dimensional invariant subspaces. Instead, the quantum eigenvalue estimation (QEVE) algorithm of Ref.~\cite{lowQuantum2024} encodes a sum of Chebyshev polynomials in $H$. Again, we start with an input state $\ket{\psi_0}$ which ideally has a high overlap with the true ground state of $H$. Instead of controlled Trotter/walk operators, we apply a circuit that prepares the \textit{Chebyshev history state}
\begin{equation}
\label{eq:history_state}
        \ket{\Phi}=\sum_{l=0}^{N-1} \ket{\ell} \, T_\ell\left(\frac{H}{\alpha}\right)\ket{\psi_0}.
\end{equation}
Here $T_l(x)$ are the Chebyshev polynomials of the first kind:
\begin{equation}
    T_\ell(x) = \cos\!\left(\ell \arccos(x)\right)
\end{equation}
The ancilla qubits required for LCU are not shown, as well as the normalization factor of $\ket{\Phi}$. 
The Chebyshev polynomials of $H / \alpha$ act on the right eigenstates as follows:
\[
    T_\ell(H/\alpha) \ket{\lambda_j} = T_\ell(\lambda_j/\alpha) \ket{\lambda_j}.
\]
In the (right) eigenstate expansion, the history state is then
\[
    \ket{\Phi} = \sum_{l=0}^{N-1} \sum_j c_j \ket{l} \, T_l(\lambda_j/\alpha) \ket{\lambda_j}
\]
After preparing $\ket{\Phi}$, we apply QFT to the ancilla register and measure it. Since $T_l(\lambda_j/\alpha) = \cos(l \theta_j)$ with $\theta_j = \arccos(\lambda_j/\alpha)$, the Fourier transform produces peaks at $\theta_j$. Measuring the ancilla yields an estimate of $\theta_j$, and hence of $\lambda_j$.

The maximum degree $N$ controls the accuracy of eigenvalue estimation: for a target accuracy $\varepsilon$, $N$ scales as 
\begin{equation}
   N =  O\left(\frac{\alpha}{\varepsilon}\right)
\end{equation}

\subsection{Constructing the Chebyshev history state using a quantum linear system solver}
\label{sec:chebyshev_theory}

Here we discuss the preparation of the history state~(\ref{eq:history_state}) for a non-Hermitian Hamiltonian $H$. Using the block encoding of $H$, one could in principle prepare the Chebyshev polynomials $T_\ell(H/\alpha)$ by preparing individual monomials of $H/\alpha$ and adding the block encodings by treating them as unitaries in another LCU expansion, but since the coefficients of the Chebyshev polynomials scale exponentially with $\ell$, so would the cost of this approach. Instead, we take advantage of the fact that we know the generating function of $T_\ell (x)$:
\begin{equation}
    G(x, y) = \sum_{\ell=0}^\infty y^\ell T_\ell(x) = \frac{1 - xy}{1 -2xy + x^2}
\end{equation}
With a lower shift operator
\begin{equation}
L := \sum_{\ell=0}^{N-2} \ket{\ell+1}\bra{\ell}    
\end{equation}
we can implement the history state
by applying the matrix version of the Chebyshev generating function:
\begin{gather}
    G(H/\alpha, L) = \sum_{\ell=0}^{\infty} L^\ell \otimes T_\ell\!\left(\frac{H}{\alpha}\right) \nonumber \\
    = \frac{1 - L \otimes H/\alpha}{1 - 2 L \otimes H/\alpha + L^2 \otimes 1}
\end{gather}
Here the series truncates at $N-1$ because $L^N=0$.
The nontrivial part then is the application of the denominator
\begin{equation}
\label{eq:denominator}
    C = 1 - 2 L \otimes H/\alpha + L^2 \otimes 1
\end{equation}
To do this, we construct a block encoding of $C$ and use a quantum linear system solver algorithm to apply $C^{-1}$. The most advanced solver that we are aware of is the adiabatic solver of Ref.~\cite{costaOptimal2022}; its runtime scales linearly with the condition number of $C$ and logarithmically with the required precision of the inversion. Internally, this algorithm constructs a Hermitian operator out of $C$ and $C^\dagger$, then implements adiabatic evolution using qubitized quantum walks.

\subsection{Resource costs}


The gate complexity and qubit count of both algorithms depend on the required accuracy of the eigenvalue estimation. We choose the total error budget of $\varepsilon=0.0016$ Hartree, an energy scale typically referred to as ``chemical accuracy''. One should keep in mind, however, that the total error of computing the ground state energy of a system consists of the error in finding the finite-basis ground state energy, as well as the basis set incompleteness error. 
Nonetheless, the choice of the error budget does not significantly alter the comparison between different methods which have the same asymptotic scaling. 

For both methods, the error budget can be split between the following sources of error. The first one is the error of the phase estimation $\varepsilon_{QPE}$. This error determines the number of ancilla qubits in QPE and the maximum Chebyshev degree in QEVE. In both cases, it ends up determining the number of times we need to apply the qubitized walk operator.

The second source of error is the error of the block encoding $\varepsilon_{BE}$, which depends on the precision with which we store the coefficients of the LCU decomposition. To bound the effect of this finite precision, we use the Weyl perturbation theorem~\cite{bhatiaMatrix1997}, which states that the eigenvalues $\lambda_j$ of a diagonalizable matrix $A = SDS^{-1}$ and the eigenvalues $\lambda'_{j'}$ of a perturbed matrix $A + E$ are connected as follows:
\begin{equation}
\label{eq:weyl}
    \max_j \min_{j'} |\lambda_j - \lambda'_{j'}| \leq ||S||\cdot ||S^{-1} || \cdot ||E||
\end{equation}
In our case, $A = H/ \alpha = \sum (b_j/\alpha) \cdot U_j$ is the Hamiltonian of interest, and $E = \sum \delta_j U_j$ is the perturbation appearing from the coefficient truncation. The value
\begin{equation}
    \kappa_S := ||S||\cdot ||S^{-1}||
\end{equation}
is also known as the Jordan condition number. Substituting these expressions into (\ref{eq:weyl}) and using the triangle inequality for the norm, we obtain
\begin{equation}
    \label{eq:weyl_2}
    \max_j \min_{j'} |\lambda_j - \lambda'_{j'}| \leq \kappa_S  K \cdot \max_j |\delta_j|
\end{equation}
The perturbation in the coefficients is bounded by $2^{-\mu}$, where $\mu$ is the number of bits of precision. The expression on the left hand side is the maximum error in the eigenvalues of $H/\alpha$ introduced by the truncation. The error bound is then the following:
\begin{equation}
\label{eq:weyl_3}
\varepsilon_{BE} \leq \alpha \kappa_S K 2^{-\mu}
\end{equation}

We now summarize the main formulas for resource estimation, while the detailed calculations are deferred to Appendices~\ref{sec:qubitization_cost} and~\ref{sec:qeve_cost}. Both for QEVE and qubitization, we assume that the runtime of the error-corrected quantum circuits is dominated by the T gates, and we ignore the cost of quantum Fourier transformation as it is much lower than all other costs considered.

\subsubsection{Qubitization} 

Ignoring the preparation of the initial state, the cost of qubitization is simply the cost of the quantum walk unitary multiplied the number of times we need to implement it (the number of ``calls'' if we treat it as a subroutine of the algorithm). The qubit budget comprises the main register storing the state of the physical system, with one qubit per each of the $2M$ spin-orbitals; the QPE register, which needs $\sim \log 1/\varepsilon_{QPE}$ qubits; the register for implementing the LCU ($\sim \log K$ qubits);and ancilla registers used in the implementation of the SELECT and PREPARE circuits. SELECT, implemented with unary iteration~\cite{babbushEncoding2018}, needs $\sim \log K$ ancilla qubits, beside the LCU register. PREPARE, implemented with coherent alias sampling~\cite{babbushEncoding2018}, needs $2 \mu + \log K + 1$ qubits, where $\mu$ is the number of bits of precision used for storing the coefficients of the LCU.

To estimate the phase of the qubitization oracle to the accuracy $\delta \varphi \sim \frac{\varepsilon_{QPE}}{\alpha}$, we need $\lceil \log \frac{2 \pi \alpha}{\varepsilon_{QPE}} \rceil + 2$ ancilla qubits and $\sim \frac{8 \pi \alpha}{\varepsilon_{QPE}}$ calls to the walk unitary. The sufficient precision of the coefficients follows from (\ref{eq:weyl_3}):
\begin{equation}
    \mu \geq \log \frac{\varepsilon_{BE}}{\alpha K},
\end{equation}
with $\kappa_S = 1$ for Hermitian matrices. In addition, if some coefficients in the block encoding end up being smaller than $2^{-\mu}$, we can discard them and get a slightly more compact block encoding. 

The cost of the quantum walk circuit mainly depends on the T gate costs of constructing the PREPARE and SELECT circuits. PREPARE is implemented using quantum read-only memory (QROM) and has the cost approximately equal to $(4 K + 4 \mu)$ T gates ~\cite{babbushEncoding2018}. SELECT is implemented using unary iteration~\cite{babbushEncoding2018} and costs $4K - 4$ T gates. Since the walk circuit needs a call to SELECT, a call to PREPARE and a call to its inverse, the total qubitization cost can be estimated as
\begin{equation}
    T_q \sim (12 K + 4 \mu) \frac{16 \pi \alpha}{\varepsilon}
\end{equation}
T gates and 
\begin{equation}
    Q_q \sim \log \frac{2 \pi \alpha}{\varepsilon_{QPE}} + 2 + 2M + \log K + 2 \mu
\end{equation}
logical qubits, assuming that we distribute the error budget equally ($\varepsilon_{QPE} = \varepsilon_{BE} = \varepsilon / 2$). Here $2M$ is the size of the main register encoding the spin-orbitals.

\subsubsection{QEVE}

The required accuracy of the phase estimation determines the required Chebyshev degree~$N$:~\cite{lowQuantum2024}
\begin{equation}
   N \sim 10 \pi \alpha / \varepsilon_{QPE}
\end{equation}
The preparation of the Chebyshev history state relies on the inversion of the denominator~(\ref{eq:denominator}). The inversion of $C$ using the method of Ref.~\cite{costaOptimal2022} requires about $4610 \sqrt{2} \kappa_C$ calls to the block encoding of $C$, where $\kappa_C$ is the condition number of $C$. The condition number of $C$ can be bounded as
\begin{equation}
    \kappa_C \leq \sqrt{24} N \kappa_S.
\end{equation}
Substituting all numbers, we find that we need 
\begin{equation*}
    184400 \cdot \sqrt{3} \pi \alpha \kappa_S / \varepsilon_{QPE}
\end{equation*}
calls to the block encoding of $C$. The block encoding of $C$ can be constructed from the block encodings of $H$ and $L$, with the former being the more expensive part. The cost of block encoding $H$ is essentially the same as in the qubitization case, $(12K + 4\mu + ...)$ T gates. The only difference is that the Jordan condition number also affects the estimate for $\mu$. In total, the QEVE algorithm requires approximately 
\begin{equation*}
    (12K + 4\mu)368800 \cdot \sqrt{3} \pi \alpha \kappa_S / \varepsilon
\end{equation*}
T gates, again assuming that we distribute the error budget equally between $\varepsilon_{QPE}$ and $\varepsilon_{BE}$.

\section{Numerical results}
\label{sec:results}

\subsection{Energy error}

The TC Hamiltonian is non-Hermitian and includes more terms than its non-TC counterpart in the same basis, so it is natural to expect the energy estimation to be more expensive. However, since the TC Hamiltonian incorporates the Jastrow factor, its ground state energy should in principle approximate the complete basis set limit more accurately. In this subsection, we compare the energy errors for second-row atoms obtained with the TC Hamiltonian in the minimal basis against those obtained with the non-TC Hamiltonian in larger basis sets. We use the experimental energies from Ref.~\cite{chakravortyGroundstate1993} as the complete basis set (CBS) limit energies. The FCI energies in the STO-6G and cc-pVDZ bases, as well as the FCI energy in the cc-pVQZ basis for Li, are calculated using PySCF~\cite{sunRecent2020}. The energies in the cc-pVTZ basis are taken from Ref.~\cite{schraivogelTranscorrelated2021}. The estimate of the FCI energy for Be in the cc-pVQZ basis was taken from the DARPA Quantum Benchmarking dataset~\cite{DarpaQB}, where it was calculated using the density matrix renormalization group (DMRG) and selected heat-bath CI (SHCI) methods.

For the TC case, the Hamiltonians were obtained using CASINO~\cite{needsVariational2020} and the TCHInt library~\cite{cohenSimilarity2019}.
We used the Drummond--Towler--Needs form of the Jastrow factor~\cite{drummondJastrow2004}. The electron-nuclear cusp was corrected using the cusp correction scheme of Refs.~\cite{maScheme2005} and~\cite{hauptOptimizing2023}.

To optimize the Jastrow factor, we used the variational Monte Carlo algorithm (VMC) with the variance of reference energy being the optimization target~\cite{hauptOptimizing2023}. The parameters of the Jastrow factor and the sampling settings were set following Refs.~\cite{hauptOptimizing2023} and~\cite{dobrautzInitio2024}. The VMC was performed in 5 cycles of sampling and optimization, with each sampling cycle returning $10^6$ points. Since the use of VMC entails randomness in the resulting Jastrow factor and consequently the TC Hamiltonian, we ran 10 independent trials for each atom.

For a single-atom system, the Drummond--Towler--Needs Jastrow factor is a sum of three terms:

\begin{gather}
    u = \sum_{ij} (r_{ij} - L_u)^3 \Theta(L_u - r_{ij}) \sum_{l = 0}^{d_u} \omega_l r_{ij}^l; \\
    \chi = \sum_{i} (r_{iI} - L_\chi)^3 \Theta(L_\chi - r_{iI}) \sum_{l=0}^{d_{\chi}} \beta_{l} r_{iI}^l\\
    f = \sum_{ij} (r_{iI} - L_f)^3 \Theta(L_f - r_{iI}) \times (r_{jI} - L_f)^3 \Theta(L_f - r_{jI}) \nonumber \\
    \times \sum_{k=0}^{d_{f, ee}} \sum_{l, m=0}^{d_{f, en}} \gamma_{klm}r_{ij}^k r_{iI}^l r_{jI}^{m}
\end{gather}

Here $L_u, L_\chi, L_f$ are (optimizable) cutoff distances, $\Theta(x)$ is the Heaviside step function, and $\omega_l, \beta_l, \gamma_{lmn}$ are optimizable parameters. For $u$, these can be different depending on whether the electrons $i$ and $j$ have the same spin or different spins. The parameters $\omega_0$ and $\omega_1$ are fixed by the electron-electron cusp condition, and the terms $\chi$ and $f$ are constrained to not affect any cusp conditions. The maximum degrees $d_u, d_\chi, d_{f,ee}, d_{f_{en}}$ control the flexibility of the Jastrow factor. For Li, Be, B, C, and N, we used the Jastrow factor with $d_u=d_\chi = 9$, $d_{f,ee}=d_{f,en} = 4$. For O, F, and Ne, VMC with the same Jastrow factor would consistently converge to obviously non-variational solutions (several Hartrees or even tens of Hartrees below the CBS energy). To mitigate this, we set $d_\chi = d_{f,ee}=d_{f,en} = 2$.

\begin{figure}
    \centering
    \includegraphics[width=\linewidth]{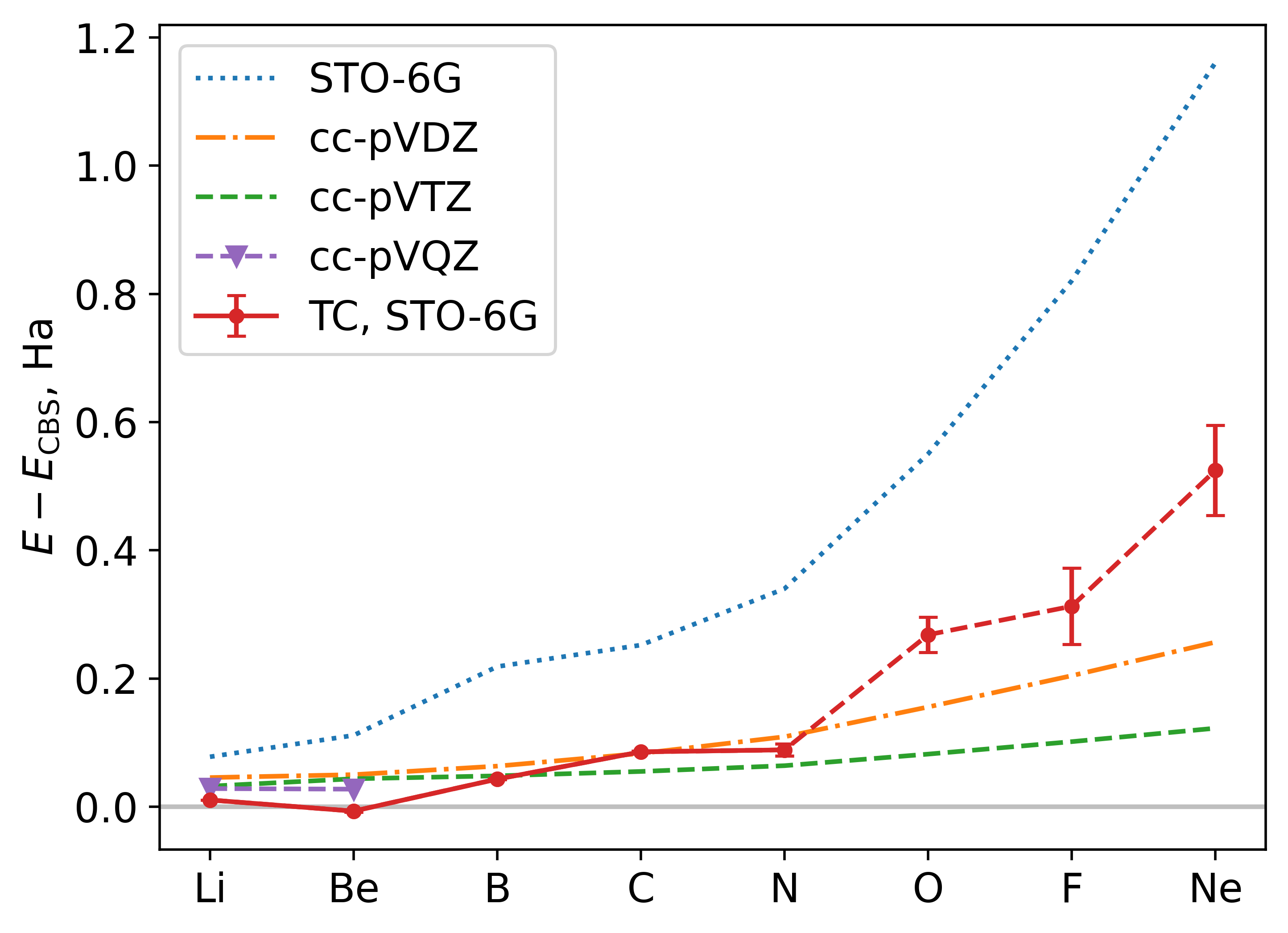}
    \caption{Energy error against the CBS limit for the transcorrelated FCI energy in the minimal basis and the non-TC FCI energy in different bases. The results for O, F, and Ne were obtained using a different Jastrow factor compared to the rest of the systems (see the main text).}
    \label{fig:energy_to_cbs}
\end{figure}

The results are shown in Figure~\ref{fig:energy_to_cbs}. For the smallest systems (Li, Be), the TC energy is substantially more accurate than the quadruple-zeta energy in the non-TC picture. For more complicated systems, the accuracy lies between that of the double- and triple-zeta bases. Finally, the TC energy for O, F, and Ne showed a larger error than the double-zeta energy.

Finally, for TC Hamiltonians, we applied the xTC approximation with the Hartree--Fock state as the reference state for the generalized normal ordering. In all cases, the energy shift introduced by xTC was 0.2 mHa or less.

Tables~\ref{tab:second_row_norms} and~\ref{tab:second_row_term_counts} show the one-norms $\alpha$ and the term counts $K$, respectively. 
By both measures, the TC Hamiltonians in the STO-6G basis are much smaller than the non-TC Hamiltonians in the cc-pVXZ bases. Moreover, even though the number of terms in the TC Hamiltonians for the STO-6G basis is several times larger than that in the non-TC, STO-6G Hamiltonians, the one-norms have the same order of magnitude. 

The randomness in the one-norms is caused by different realizations of the VMC calculation. The number of terms is also random because for the exact TC Hamiltonian in the second-quantized form, we only keep the terms with absolute values of coefficients above $10^{-15}$. We further discard terms according to the threshold that we obtain from (\ref{eq:weyl_3}) when distributing the error budget (denoted as ``Truncated TC'' in Table~\ref{tab:second_row_term_counts}). In all cases, the threshold was rather small (around $10^{-8}$ to $10^{-7}$ depending on the atom and the instance of VMC calculation), so discarding the terms with weights smaller than that threshold did not affect the one-norms or energies in a noticeable way (the energy changed by $10^{-7}$ Hartree or less). The xTC Hamiltonians did not have any terms that could be discarded this way.

\begin{table}
    \centering
    \resizebox{\columnwidth}{!}
    {\begin{tabular}{|c|c|c|c|c|c|c|c|c|}
    \hline
          & Li & Be & B & C & N & O & F & Ne \\
          \hline
         STO-6G & 	8.2 & 12.0 & 	17.4 & 	26.0 & 	36.2 & 48.4 & 62.3 & 	78.2 \\
        cc-pVDZ & 67.4 & 112.5 & 121.7 & 154.5 & 202.8 & 203.6 & 242.0	& 293.7 \\
        cc-pVTZ & $5.83 \times 10^2$ & $8.39 \times 10^2$ & $8.91 \times 10^2$ & $1.10 \times 10^3$ &$1.63 \times 10^3$ &$1.53 \times 10^3$  &$1.76 \times 10^3$	&$2.39 \times 10^3$ \\
        cc-pVQZ & $2.80 \times 10^3$ & $4.32 \times 10^3$ & $4.00 \times 10^3$ & $5.20 \times 10^3$ & $8.17 \times 10^3$ & $7.00 \times 10^3$ & $8.06 \times 10^3$ & $1.20 \times 10^4$ \\
        TC, STO-6G & $8.298 \pm 0.008$ &
    $12.187 \pm 0.010$ &
    $18.155 \pm 0.006$ &
    $26.939 \pm 0.010$ &
    $37.709 \pm 0.010$ &
    $50.367 \pm 0.006$ &
    $65.068 \pm 0.009$ &
    $81.784 \pm 0.097$  \\
        xTC, STO-6G & $7.939 \pm 0.009$ &
    $12.004 \pm 0.009$ &
    $18.100 \pm 0.006$ &
    $26.890 \pm 0.010$ &
    $37.630 \pm 0.011$ &
    $50.231 \pm 0.021$ &
    $64.778 \pm 0.038$ &
    $81.416 \pm 0.180$  \\
        \hline
    \end{tabular}}
    \caption{One-norms of second-row atoms in their Pauli decompositions.}
    \label{tab:second_row_norms}
\end{table}

\begin{table*}
    \centering
    \resizebox{\columnwidth}{!}
    {\begin{tabular}{|c|c|c|c|c|c|c|c|c|}
    \hline
         & Li & Be & B & C & N & O & F & Ne \\
         \hline
        STO-6G & 154 & 154& 154& 154& 154& 154& 154& 154 \\        
        cc-pVDZ & $1.27 \times 10^4$ & $2.25 \times 10^4$ & $1.10 \times 10^4$ & $1.10 \times 10^4$  & $2.30 \times 10^4$ & $1.10 \times 10^4$ & $1.10 \times 10^4$ & $2.30 \times 10^4$  \\
        cc-pVTZ & $5.02 \times 10^5$ & $5.37 \times 10^5$  & 	$2.39 \times 10^5$  &	$2.39 \times 10^5$  & $5.47 \times 10^5$  &	$2.39 \times 10^5$  &	$2.39 \times 10^5$  &	$5.48 \times 10^5$ \\
        cc-pVQZ & $6.10 \times 10^6$ 	& $6.23 \times 10^6$  & $2.62 \times 10^6$	 & $2.62 \times 10^6$  & $6.39 \times 10^6$	 &$2.61 \times 10^6$ &$2.62 \times 10^6$ &$6.39 \times 10^6$ \\
        TC, STO-6G & 
        $1088.6 \pm 62.9$ &
        $965.0 \pm 51.6$ &
        $959.0 \pm 0.0$ &
        $930.2 \pm 52.8$ &
        $969.0 \pm 32.0$ &
        $997.4 \pm 34.8$ &
        $953.7 \pm 10.6$ &
        $959.8 \pm 70.5$ 
        \\
        Truncated TC & 
        $978.2 \pm 51.6$ &
        $944.6 \pm 23.2$ &
        $959.0 \pm 0.0$ &
        $915.8 \pm 15.2$ &
        $956.6 \pm 7.6$ &
        $956.6 \pm 5.1$ &
        $953.7 \pm 10.6$ &
        $933.8 \pm 75.6$
        \\
        xTC, STO-6G & $295.0 \pm 0.0$ &
$295.0 \pm 0.0$ &
$295.0 \pm 0.0$ &
$295.0 \pm 0.0$ &
$295.0 \pm 0.0$ &
$295.0 \pm 0.0$ &
$295.0 \pm 0.0$ &
$295.0 \pm 0.0$  \\
        \hline
    \end{tabular}}
    \caption{Number of terms for second-row atoms in the Pauli decompositions. Error intervals denote one standard deviation.}
    \label{tab:second_row_term_counts}
\end{table*}

\begin{figure}
    \centering
    \includegraphics[width=\linewidth]{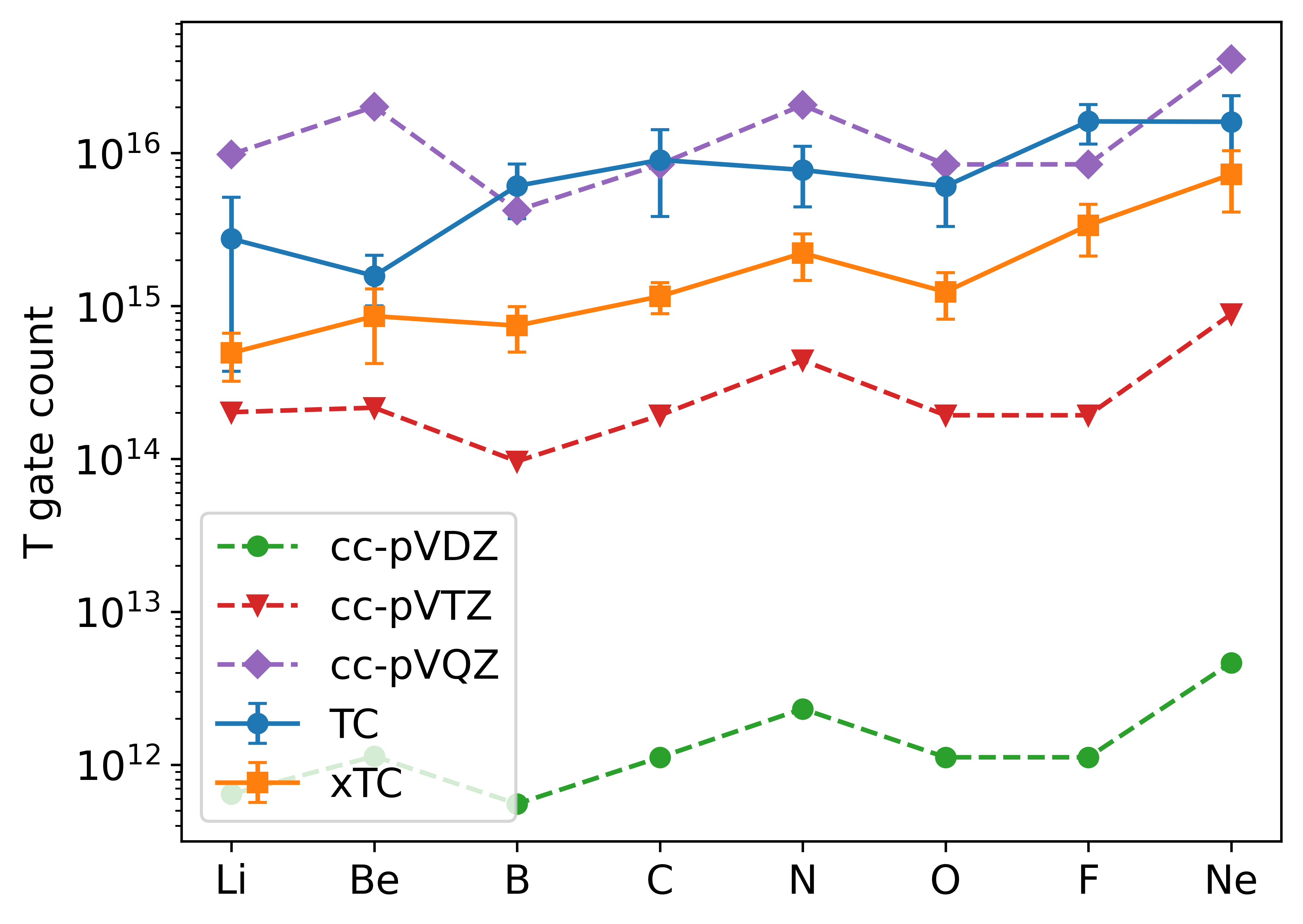}
    \caption{T gate count estimates for QEVE (solid lines) and qubitization (dashed lines).}
    \label{fig:t_counts}
\end{figure}

Figure~\ref{fig:t_counts} shows the gate counts for performing the qubitization and QEVE for the second-row atoms. While the exact TC method shows gate counts comparable to those of the cc-pVQZ method, the xTC approximation significantly reduces the gate counts, although they still remain above the cc-pVTZ level.

The qubit requirements did not change significantly across different systems. On average, QEVE requires 127.8 logical qubits for the TC Hamiltonians and 118.3 qubits for the xTC Hamiltonians, while standard qubitization requires on average 172.2, 240.2, and 318.1 qubits for cc-pVDZ, cc-pVTZ, and cc-pVQZ bases respectively. 

For both algorithms, there is some flexibility in implementing the PREPARE subroutine. The QROAM algorithm~\cite{berryQubitization2019} allows one to reduce the number of T gates at the cost of introducing more ancilla qubits. In Appendix~\ref{sec:qroam}, we show the T gate and qubit counts for the case when QROAM is optimized for the gate count. In this case, the advantage in gate count is limited, but the qubit count is drastically increased, making this approach impractical.

\begin{figure}
    \centering
    \includegraphics[width=\linewidth]{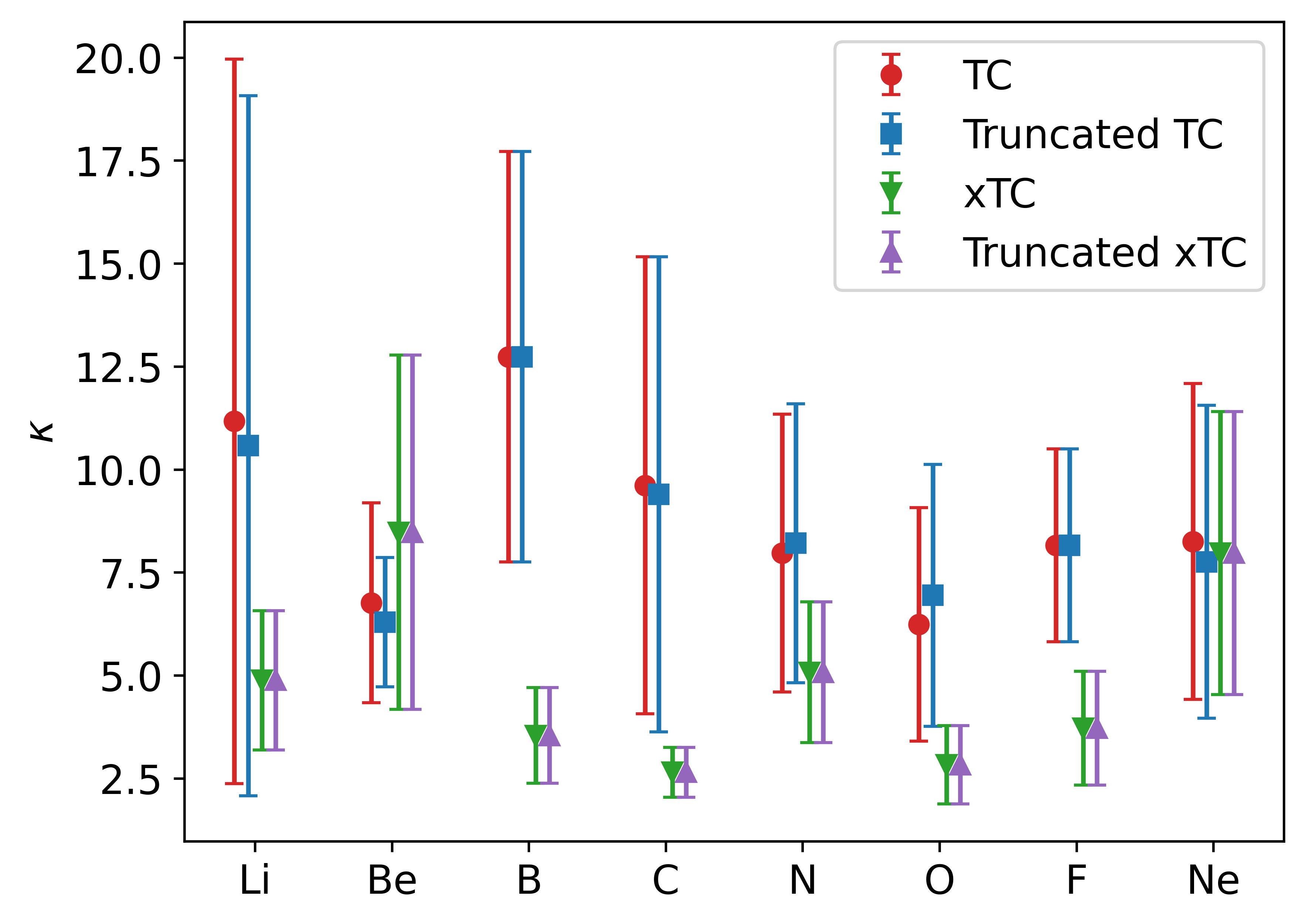}
    \caption{Jordan condition numbers for the TC and xTC Hamiltonians.}
    \label{fig:conds}
\end{figure}

Figure~\ref{fig:conds} shows the Jordan condition number of the qubit Hamiltonians across the systems. The truncation of the Hamiltonian that we used for QEVE does not significantly change the condition number. However, the xTC approximation reduces it quite significantly in almost all instances. We calculated $\kappa_S$ via exact diagonalization of the Hamiltonian in qubit form: it equals the condition number of the matrix of normalized right eigenvectors.

\section{Conclusions}
\label{sec:conclusions}

The transcorrelated method significantly reduces the basis set error. However, the computational overhead stemming from the non-Hermiticity of the TC Hamiltonian is hard to outweigh, and turning to the transcorrelated picture does not always lead to better quantum gate counts. The reason for the overhead is rooted in the inversion algorithm of Costa et al.~\cite{costaOptimal2022}. Despite leading to optimal asymptotic scaling, this algorithm also contributes a large constant factor to the overall complexity.

Estimating the Jordan condition number $\kappa_S$ remains a challenge. Even though the similarity transformation in first quantization is generated by a well-behaved operator, there is no guarantee that the projected Hamiltonian can be diagonalized by a similarity transformation with a small condition number. Understanding the behavior of $\kappa_S$ is critical for determining the success of QEVE and potentially any other quantum algorithm that deals with the spectrum of diagonalizable non-Hermitian matrices. Since the non-normality of $H$ is caused by the Jastrow factor, one could estimate the condition number as
\begin{equation}
    \kappa_S \sim ||e^{\hat{J}}|| \cdot ||e^{-\hat{J}}||
\end{equation}
However, this estimate is hard to justify rigorously, since it is not clear how exactly the non-normality is carried through the projection onto a finite-dimensional subspace.

In our experiments, we observed that the TC energy was sensitive to the choice of the Jastrow factor. Choosing its form and optimization routine more carefully could potentially improve the energy estimates. For example, a recent work by Filip et al.~\cite{filipDeterministic2025} showed that one can achieve accurate results using deterministically optimized Jastrow factors. 

The performance of all QPE-based algorithms for ground state energy estimation depends on the overlap between the input state $\ket{\psi_0}$ and the true ground state. Multiple works on the transcorrelated method have reported that the ground state of a TC Hamiltonian admits a more compact representation than that of a non-TC Hamiltonian~\cite{dobrautzCompact2019,liaoefficient2021,schraivogelTranscorrelated2021,liaoDensity2023}. This compactness may make the TC treatment more favorable.

The Hamiltonians used in the work, the input CASINO files, and the code for resource estimation are available on Github: \href{https://github.com/aleksey-uvarov/qeve_cost}{https://github.com/aleksey-uvarov/qeve\_cost}.

\section{Acknowledgements}

We thank Nathan Wiebe for helpful discussions. A.U.~thanks Pablo L\'opez R\'ios and Werner Dobrautz for their help in configuring TCHInt. This work was supported by Mitacs through the Mitacs Elevate program. This research was partly enabled by the support of Compute Ontario (computeontario.ca) and the Digital Research Alliance of Canada (alliancecan.ca). Part of the computations were performed on the Niagara and Trillium supercomputers at the SciNet HPC Consortium, and the NARVAL and RORQUAL supercomputers under the Calcul Quebec Consortium. SciNet is funded by Innovation, Science, and Economic Development Canada, the Digital Research Alliance of Canada, the Ontario Research Fund: Research Excellence, and the University of Toronto.

\appendix

\section{Complexity analysis for qubitization}
\label{sec:qubitization_cost}

\subsection{Query complexity}

A key component of qubitization is the controlled walk operator. Its preparation requires one call to SELECT, one call to PREPARE, one call to $\mathrm{PREPARE}^\dagger$, and one call to a multi-controlled Z. The target eigenvalue precision determines the number of times the controlled walk operator $W$ must be applied. Finally, we apply the quantum Fourier transform to the ancilla register and measure.

The spectrum of the walk operator

$$
W = (2 \ket{0} \bra{0} - 1) \cdot \mathrm{PREP}^\dagger \cdot 
\mathrm{SEL} \cdot \mathrm{PREP}
$$

consists of values of the form $\exp(\pm i \arccos \frac{\lambda_k}{\alpha})$. The phase estimation algorithm will extract the phase $\varphi = \arccos \frac{\lambda_k}{\alpha}$ up to some precision $\delta \varphi$. We want this phase error to be such that its contribution to the total energy error is at most $\varepsilon_{QPE}$. By considering the derivative of the arccosine, we can estimate that at worst we need $\delta \varphi \approx \frac{\varepsilon_{QPE}}{\alpha}$.

To perform phase estimation successfully with probability at least $(1 - p_{fail})$, we need
\begin{equation}
\label{eq:qpe_qubits}
    n_a = \left\lceil \log \frac{2 \pi}{ \delta \varphi}\right\rceil + \left\lceil \log (2 + \frac{1}{2 p_{fail}})\right\rceil 
\end{equation}
ancilla qubits~\cite{nielsenQuantumComputationQuantum2010}. Thus, we will need $2^{n_a} - 1$ controlled applications of $Q$. If we demand $p_{fail} < 1/2$, the second term in~(\ref{eq:qpe_qubits}) yields two extra qubits. The number of calls to the walk circuit is then
\begin{equation}
    2^{n_a} - 1 = 2^{\lceil \log \frac{2 \pi \alpha}{ \varepsilon_{QPE}} \rceil + 2} - 1 = 4 \cdot 2^{\lceil \log \frac{2 \pi \alpha}{ \varepsilon_{QPE}}\rceil} -1.
\end{equation}

\subsection{The cost of PREPARE}

We follow the realization of PREPARE given in Ref.~\cite{babbushEncoding2018}. In short, it first uses QROM (or QROAM~\cite{berryQubitization2019}) to prepare a state
\begin{equation*}
    \frac{1}{\sqrt{K}}\sum_{i=0}^{K-1} \ket{i} \ket{bin(b_{i+1})},
\end{equation*}
where $\ket{bin(b_i)}$ stores the first $\mu$ bits of the binary expansion of the coefficient $b_i$. Then using a technique called coherent alias sampling, this state is mapped to a state proportional to
\begin{equation}
    \sum_{i=0}^{K-1} \sqrt{b_{i+1}} \ket{i} \ket{\mathrm{garbage}_i}
\end{equation}
where $\ket{\mathrm{garbage}_i}$ denotes the state of ancilla qubits. The procedure includes the following steps:
\begin{enumerate}
    \item Prepare a uniform superposition
    \begin{equation}
        \frac{1}{\sqrt{K}}\sum_{i=1}^K \ket{i}
    \end{equation}
    When $K$ is a power of two, this step consists of $ \log K $ Hadamard gates, but otherwise requires a more complicated circuit.
    \item Prepare a state
    \begin{equation}
        \frac{1}{\sqrt{K}}\sum_{i=0}^{K-1} \ket{i} \ket{\mathrm{alt}_i} \ket{\mathrm{keep}_i}
    \end{equation}
    Here the registers contain $\lceil \log K \rceil$, $\lceil \log K \rceil$, and $\mu$ qubits, respectively. This step is performed using QROAM~\cite{berryQubitization2019}. We are free to select an integer $q$ which is a power of two and $1 \leq q < K$. This step will then cost 
    \begin{equation}
    \left\lceil \frac{K}{q} \right\rceil + (\mu + \lceil \log K \rceil) (q - 1)
    \end{equation}
    Toffoli gates and $(\mu + \lceil \log K \rceil) (q - 1) + \lceil \log K / q \rceil$ ancillas. Here $q=1$ corresponds to standard QROM, and increasing $q$ lets one decrease the T gate count at the cost of using more ancilla qubits.
    \item Prepare a uniform superposition in another $\mu$-qubit register $\ket{\sigma}$, then swap the index and ``alt'' register controlled on whether the value of $\sigma$ is smaller than $\mathrm{keep}_i$. This means that we need to perform an addition ($4 \mu$ T gates) and a controlled SWAP. Since a SWAP can be implemented with three CNOTs, a cSWAP can be implemented with three Toffolis. Thus we have $3 \lceil \log K \rceil$ Toffoli gates. The comparison needs one more ancilla qubit.
\end{enumerate}
The values of $\mathrm{keep}_i$ and $\mathrm{alt}_i$ can be classically engineered so that in the end we obtain a state proportional to
\begin{equation}
    \sum_{i=0}^{K-1} \sqrt{b_i} \ket{i} \ket{\mathrm{garbage}_i},
\end{equation}
as desired. 
When $K$ is not a power of two, we can nonetheless prepare a uniform distribution on the $\lceil \log K \rceil$ qubits. The desired distribution is then reachable in the same fashion.

Assuming 1 Toffoli is equal to 4 T gates, the cost of PREPARE with this procedure is
\begin{equation}
    4\left\lceil \frac{K}{q} \right\rceil + 4(\mu + \lceil \log K \rceil) (q - 1) + 4 \mu + 12 \lceil \log K \rceil
\end{equation}
T gates. Here $1 \leq q < K $ is a power of two which can be chosen in the QROAM subroutine. $q=1$ corresponds to standard QROM, and increasing $q$ lets one decrease the T gate count at the cost of using more ancilla qubits.

The total amount of ancilla qubits is 
\begin{equation}
    (\mu + \lceil \log K \rceil )(q-1) + 2 \mu + \lceil \log K \rceil + \lceil \log K / q \rceil + 1
\end{equation}

While the inverse of $\mathrm{PREP}$ could potentially be made cheaper than $\mathrm{PREP}$, here we assume that the inverse costs the same.

\subsection{The cost of SEL}

The SELECT circuit costs $4K - 4$ T gates due to unary iteration plus the cost of all controlled applications of the unitaries themselves. If these are controlled Paulis, then their T-cost is zero. The SELECT circuit, implemented with unary iteration, requires $\lceil \log K \rceil$ ancilla qubits.

\subsection{The cost of reflection}

The reflection operator is an $\lceil \log K \rceil$-controlled Z, which is equivalent to an $\lceil \log K \rceil$-Toffoli gate up to Clifford gates. This costs $4 (\lceil \log K \rceil - 1)$ T gates~\cite{jonesLowoverheadConstructionsFaulttolerant2013}.

\subsection{Total cost of a quantum walk oracle}

Putting it all together, one call of a controlled quantum walk oracle costs
\begin{equation}
    \label{eq:qubitization_cost_incomplete}
    4 K + 8 \mu + 8 \left(q - 1\right) \left(\mu + \left\lceil{\log K}\right\rceil\right) + 8 \left\lceil{\frac{K}{q}}\right\rceil + 28 \left\lceil \log K \right\rceil - 8
\end{equation}
T gates. The number of ancilla qubits required is 
\begin{equation}
    (\mu + \lceil \log K \rceil )(q-1) + 2 \mu + 2 \lceil \log K \rceil + \lceil \log K / q \rceil + 1
\end{equation}

\section{Complexity analysis for QEVE}
\label{sec:qeve_cost}

By $N - 1$ we denote the maximum degree of the Chebyshev polynomials, and by $n = \log N$ the corresponding number of ancilla qubits. We assume that $||H/\alpha|| \leq 1/2$, this is necessary for the QEVE algorithm and can be ensured at no extra cost in T gates~\cite{lowQuantum2024}. We also assume that $N$ is a power of two (for cost estimations, when we calculate the value of $N$ needed to reach the required accuracy, we round it up to the nearest power of two).

\subsection{Block encoding of the denominator}

The QEVE algorithm uses a block encoding of $C = (1 + L^2 \otimes 1 - 2 L \otimes H/\alpha)$. This block encoding needs block encodings of $H/\alpha$ and $L$. The block encoding of $H$ has the same T cost as for the qubitization case, namely the cost of SELECT plus twice the cost of PREPARE:
\begin{equation}
    \mathcal{C} = 4 K + 8 \mu + 8 \left(q - 1\right) \left(\mu + \left\lceil{\log K}\right\rceil\right) + 8 \left\lceil \frac{K}{q}  \right\rceil + 24 \left\lceil{\log K}\right\rceil - 4
\end{equation}

The lower shift $L$ is implemented using the cyclic shift operator
\begin{equation}
X_{2N} = \sum_{j = 0}^{2N-1} \ket{j + 1 \operatorname{mod} 2N}\bra{j}   
\end{equation}
together with a \emph{comparator} gate to filter out the unnecessary blocks:
\begin{equation}
    CMP_{2N} = \sum_{k\leq N-1} |k \rangle \langle k| \otimes I + \sum_{k>{N-1}} |k \rangle \langle k| \otimes X
\end{equation}
When $N$ is a power of two, this gate is just a CNOT. The block encoding of $L$ requires two comparators and one shift. 

We can build the shift operator by composing an $(n-1)$-Toffoli gate, an $(n-2)$-Toffoli, $(n-3)$-Toffoli, and so on (see e.g.~Ref.~\cite{liCLASS2014}, Figure 4). The cost of this implementation is $4 (n-2) + 4(n-3) + ... + 4 = 2(n-1)(n-2)$ T gates. The controlled implementation of the shift operator means adding one more control to every Toffoli, making the overall cost equal to $2n(n-1)$ T gates.

We can encode $C$ using the block encodings of $L$ and $H/\alpha$ as LCU components. The one-norm of this LCU is 4, so we get a block encoding of $C / 4$. The cost is $\mathcal{C}_{BE(H)} + 6n(n-1)$. The PREPARE circuit of this LCU doesn't cost any extra T gates: we need two ancilla qubits, and we can rewrite the LCU as $1 + L^2 - LH - LH$. This way all coefficients are equal to one up to phase, and the PREPARE circuit is just two Hadamard gates.

The operator norm of $C/4$ is less than or equal to 3/4, which can be shown directly using the properties of the norm (the norm of $L$ is equal to one: $L L^\dagger$ is diagonal and has eigenvalues 0 and 1). 
On the other hand, by directly applying $C$ to a vector $\ket{0} \ket{\psi}$, we can find that $||C|| \geq \sqrt{2+ 4 ||H||^2} \geq \sqrt 2$.

\subsection{Quantum linear system solver}

The core of the QEVE algorithm is the quantum linear systems solver algorithm of Ref.~\cite{costaOptimal2022} (or any algorithm for solving linear systems). Given $C$ and $\ket{b}$ as an input, it prepares a state proportional to $C^{-1} \ket{b}$. It consists of two steps: adiabatic evolution and eigenstate filtering. They both scale linearly with the condition number of $C$, but the filtering step also carries a logarithmic dependence on the accuracy of preparing the solution. However, compared to the constant factor appearing in the adiabatic evolution step, the filtering step has a negligible cost for any reasonable accuracy requirements. 

In the adiabatic evolution step, the algorithm uses a qubitized quantum walk of the Hermitian operator
\begin{equation}
\label{eq:walker_hamiltonian}
    \mathbf{C} =\begin{pmatrix} 0 & C^\dagger \\
        C & 0
    \end{pmatrix} 
    = \sum_{i} c_i
    \begin{pmatrix} 0 & U_i^\dagger \\
        U_i & 0
    \end{pmatrix} 
\end{equation}
Since the Hamiltonians we consider are real, all Pauli coefficients are either purely real (in which case a Pauli string $U_i$ contains an even number of $Y$ Pauli matrices) or purely imaginary (an odd number of $Y$ Pauli matrices), and by convention that $c_i$ are real, in this case $U_i$ is also multiplied by $i$. In either case, the summands in (\ref{eq:walker_hamiltonian}) are obtained from $U_i$ by tensor multiplying them either by $X$ or by $Y$. Therefore, block encoding $\mathbf{C}$ takes the same amount of T gates as encoding $C$. To make a walk operator, we also need a reflection, which, taking all extra ancillas into account, takes $4(\lceil \log K \rceil+2 -1)$ T gates. A parametrized walk operator based on $\mathbf{C}$ is then run $T$ times to approximately implement a unitary evolution $U(s), s \in [0, 1]$.

Aside from a much looser rigorous bound, Ref.~\cite{costaOptimal2022} gives the following bound based on numerical experiments:
\begin{equation}
    ||U(s) - U_C(s)||\leq 2305 \frac{\kappa}{T}
\end{equation}
Here $\kappa$ is the condition number of the input matrix, and $U_C(s)$ is the approximate evolution. For the algorithm to work, the error of the evolution after this process has to be such that the overlap with the ground state of the final matrix is no less than 1/2. This implies that $T$ has to be chosen so that 
\begin{equation}
    \frac{1}{\sqrt 2} \geq 2305 \frac{\kappa}{T}
\end{equation}
That is,
\begin{equation}
    T \geq 2305 \kappa \sqrt2 
\end{equation}
Since the probability of success in this case is about one half, this means that on average we will need
\begin{equation}
    4610 \kappa \sqrt 2
\end{equation}
calls to the block encoding of $C$.

The condition number $\kappa$ is defined as the following product of norms:

\begin{equation}
    \kappa = ||C||\cdot ||C^{-1}||
\end{equation}

For Lemma 1 of Ref.~\cite{lowQuantum2024} we know an upper bound:
\begin{equation}
\label{eq:inverse_bound_U}
    ||C^{-1}|| \leq N \alpha_\mathrm{U} 
\end{equation}
Here
\begin{equation}
    \alpha_{\mathrm{U}} \geq \max_{0 \leq j \leq n-1} ||U_j(H / \alpha)|| 
\end{equation}
is an upper bound on the values of Chebyshev polynomials of the second kind. If we assume that $H$ is diagonalizable, we have the following bound:
\begin{equation}
\label{eq:alpha_u_bound}
    \alpha_{\mathrm{U}} \leq O(\kappa_S)
\end{equation} 
If $H = SDS^{-1}$, we can rewrite $U_j(H)$ as $SU_j(D) S^{-1}$.

To turn the ``O-large'' bound~(\ref{eq:alpha_u_bound}) into a concrete bound, we use a relation connecting the Chebyshev polynomials of first and second kind:
\begin{equation}
    T_n^2(x) - (x^2 - 1) U_{n-1}^2(x) = 1
\end{equation}
For $x \in [-1/2, 1/2]$, it follows that $|U(x)| \leq \sqrt{\frac83}$. This way, we obtain 
\begin{equation}
    \alpha_{\mathrm{U}} \leq \sqrt{\frac83} \kappa_S
\end{equation}
Collecting all the terms, we find that we need
\begin{equation*}
    18440 \cdot \sqrt{3} N \kappa_S
\end{equation*}
calls to the walk operator.

The required degree of the Chebyshev polynomial depends on the accuracy we need. In QEVE, the energy is estimated as

\begin{equation}
    \alpha \cos 2 \pi \phi
\end{equation}

where $\phi \in [1/6, 1/3]$ is measured in the device (the angle is guaranteed to be in this range by $\alpha \geq ||H||/2$). The angle is measured up to $1/N$. Ref.~\cite{lowQuantum2024} shows that, when the correct angle is not exactly an integer multiple of $1/N$, the results of the measurement are spread in such a way that the measured angle is within $5/N$ of the correct one more than half the time. Using that, if we demand to measure the energy up to $\varepsilon_{QEVE}$, we can estimate the required Chebyshev degree as

\begin{equation}
   N \sim 10 \pi \alpha / \varepsilon_{QEVE}
\end{equation}
The call count is then
\begin{equation*}
    184400 \cdot \sqrt{3} \pi \alpha \kappa_S / \varepsilon_{QEVE}
\end{equation*}




\section{Resource estimates with QROAM}\label{sec:qroam}

Using the QROAM subroutine~\cite{berryQubitization2019} of the PREPARE circuit, one can decrease the T gate count at the cost of introducing additional ancilla registers. 

\begin{figure}
    \centering
    \includegraphics[width=\linewidth]{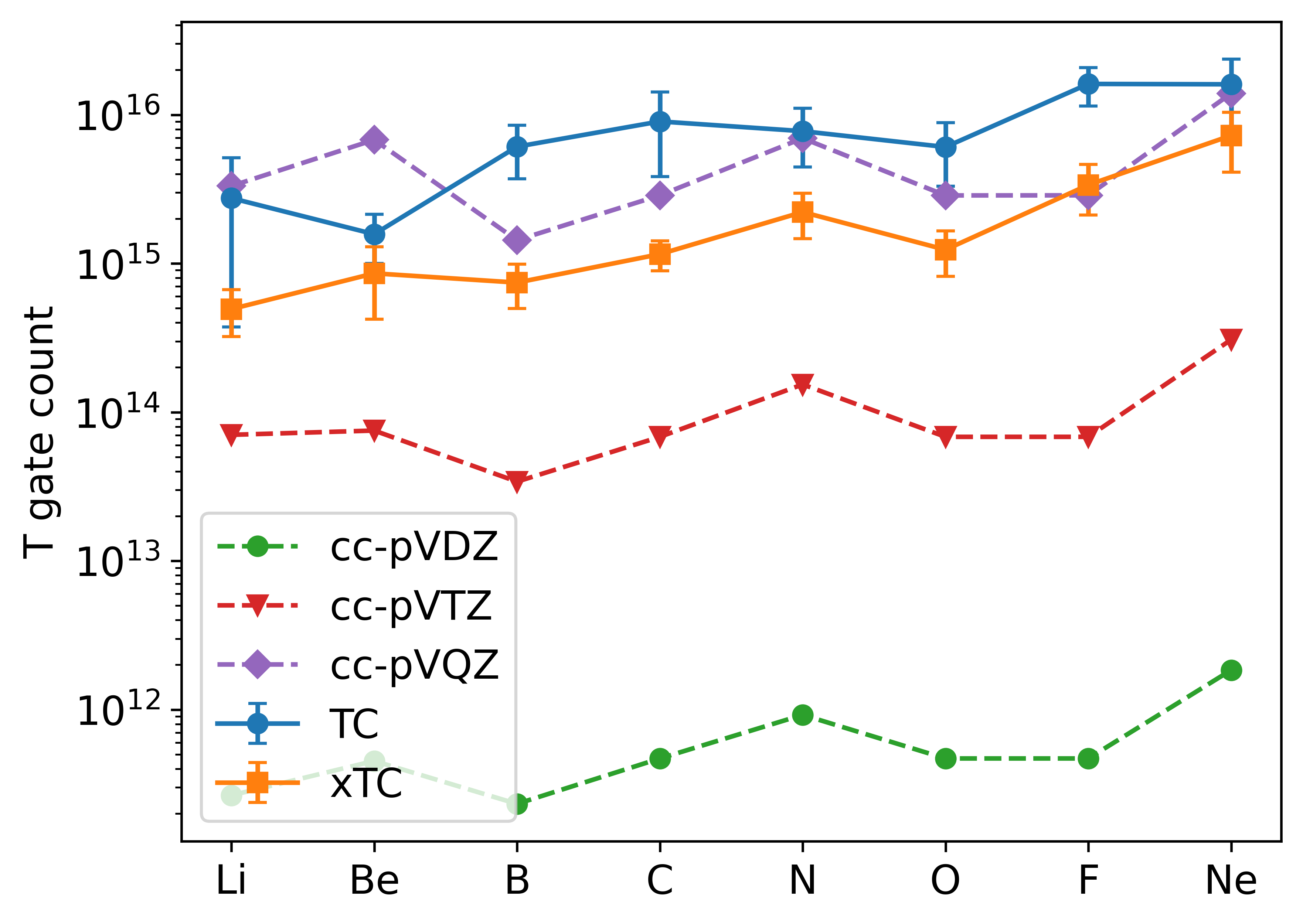}
    \caption{T gate counts obtained with an optimized QROAM subroutine.}
    \label{fig:t_counts_qroam}
\end{figure}

Figure~\ref{fig:t_counts_qroam} shows the optimized T gate counts. Optimizing the QROAM subroutine yields an improvement in T gate count up to about a factor of three, especially for the VQZ basis, but the qubit overhead quickly becomes overwhelming: 297.2 qubits on average for TC-QEVE and 255.8 qubits for xTC-QEVE; 1128.3, 7786.8, and 23323.9 qubits on average for qubitization in the cc-pVDZ, cc-pVTZ, and cc-pVQZ bases respectively. The reason for this is that QROAM does not affect the cost of the SELECT circuit, so after a certain point it becomes the most expensive part of the quantum walk, and further improvements to PREPARE become unimportant compared to the cost of SELECT.

\bibliography{apssamp}

@article{maScheme2005,
  title = {Scheme for Adding Electron--Nucleus Cusps to {{Gaussian}} Orbitals},
  author = {Ma, A. and Towler, M. D. and Drummond, N. D. and Needs, R. J.},
  year = 2005,
  month = jun,
  journal = {The Journal of Chemical Physics},
  volume = {122},
  number = {22},
  pages = {224322},
  issn = {0021-9606, 1089-7690},
  doi = {10.1063/1.1940588},
  urldate = {2026-02-04},
  langid = {english}
}

@article{hauptOptimizing2023,
  title = {Optimizing {{Jastrow}} Factors for the Transcorrelated Method},
  author = {Haupt, J. Philip and Hosseini, Seyed Mohammadreza and L{\'o}pez R{\'i}os, Pablo and Dobrautz, Werner and Cohen, Aron and Alavi, Ali},
  year = 2023,
  month = jun,
  journal = {The Journal of Chemical Physics},
  volume = {158},
  number = {22},
  pages = {224105},
  issn = {0021-9606, 1089-7690},
  doi = {10.1063/5.0147877},
  urldate = {2026-01-28},
  langid = {english}
}

@article{cohenSimilarity2019,
  title = {Similarity Transformation of the Electronic {{Schr\"odinger}} Equation via {{Jastrow}} Factorization},
  author = {Cohen, Aron J. and Luo, Hongjun and Guther, Kai and Dobrautz, Werner and Tew, David P. and Alavi, Ali},
  year = 2019,
  month = aug,
  journal = {The Journal of Chemical Physics},
  volume = {151},
  number = {6},
  pages = {061101},
  issn = {0021-9606, 1089-7690},
  doi = {10.1063/1.5116024},
  urldate = {2024-10-08},
  langid = {english}
}

@article{packCusp1966,
  title = {Cusp {{Conditions}} for {{Molecular Wavefunctions}}},
  author = {Pack, Russell T and Brown, W. Byers},
  year = 1966,
  month = jul,
  journal = {The Journal of Chemical Physics},
  volume = {45},
  number = {2},
  pages = {556--559},
  issn = {0021-9606, 1089-7690},
  doi = {10.1063/1.1727605},
  urldate = {2025-11-06},
  langid = {english}
}

@article{xieVariational2024,
  title = {Variational Quantum Algorithms for Scanning the Complex Spectrum of Non-{{Hermitian}} Systems},
  author = {Xie, Xu-Dan and Xue, Zheng-Yuan and Zhang, Dan-Bo},
  year = 2024,
  month = aug,
  journal = {Frontiers of Physics},
  volume = {19},
  number = {4},
  eprint = {2305.19807},
  primaryclass = {quant-ph},
  pages = {41202},
  issn = {2095-0462, 2095-0470},
  doi = {10.1007/s11467-023-1382-3},
  urldate = {2025-11-04},
  archiveprefix = {arXiv},
  langid = {english}
}

@article{solinasBiorthogonal2025,
  title = {Biorthogonal {{Neural Network Approach}} to {{Two-Dimensional Non-Hermitian Systems}}},
  author = {Solinas, Massimo and Barton, Brandon and Zhang, Yuxuan and Nys, Jannes and Carrasquilla, Juan},
  year = 2025,
  month = aug,
  pages = {arXiv:2508.01072},
  eprint = {2508.01072},
  primaryclass = {quant-ph},
  publisher = {arXiv},
  doi = {10.48550/arXiv.2508.01072},
  urldate = {2025-08-27},
  archiveprefix = {arXiv},
  langid = {english},
  journal={arXiv}
}

@article{peruzzoVariationalEigenvalueSolver2014,
  title = {A Variational Eigenvalue Solver on a Photonic Quantum Processor},
  author = {Peruzzo, Alberto and McClean, Jarrod and Shadbolt, Peter and Yung, Man-Hong and Zhou, Xiao-Qi and Love, Peter J. and {Aspuru-Guzik}, Al{\'a}n and O'Brien, Jeremy L.},
  year = 2014,
  month = dec,
  journal = {Nature Communications},
  volume = {5},
  number = {1},
  pages = {4213},
  issn = {2041-1723},
  doi = {10.1038/ncomms5213},
  urldate = {2018-12-11},
  langid = {english}
}

@article{yanaiCanonical2012,
  title = {Canonical Transcorrelated Theory with Projected {{Slater-type}} Geminals},
  author = {Yanai, Takeshi and Shiozaki, Toru},
  year = 2012,
  month = feb,
  journal = {The Journal of Chemical Physics},
  volume = {136},
  number = {8},
  pages = {084107},
  issn = {0021-9606, 1089-7690},
  doi = {10.1063/1.3688225},
  urldate = {2025-11-04},
  langid = {english}
}

@book{helgakerMolecular2014,
  title = {Molecular {{Electronic-Structure Theory}}},
  author = {Helgaker, Trygve and J{\o}rgensen, Poul and Olsen, Jeppe},
  year = 2014,
  publisher = {Wiley},
  address = {Hoboken},
  isbn = {978-1-118-53147-1},
  langid = {english}
}

@article{lowHamiltonian2019,
  title = {Hamiltonian {{Simulation}} by {{Qubitization}}},
  author = {Low, Guang Hao and Chuang, Isaac L.},
  year = 2019,
  month = jul,
  journal = {Quantum},
  volume = {3},
  eprint = {1610.06546},
  primaryclass = {quant-ph},
  pages = {163},
  issn = {2521-327X},
  doi = {10.22331/q-2019-07-12-163},
  urldate = {2023-03-31},
  archiveprefix = {arXiv},
  langid = {english}
}

@article{katoEigenfunctions1957,
  title = {On the Eigenfunctions of Many-particle Systems in Quantum Mechanics},
  author = {Kato, Tosio},
  year = 1957,
  month = jan,
  journal = {Communications on Pure and Applied Mathematics},
  volume = {10},
  number = {2},
  pages = {151--177},
  issn = {0010-3640, 1097-0312},
  doi = {10.1002/cpa.3160100201},
  urldate = {2025-01-07},
  langid = {english}
}

@article{whitfieldSimulationElectronicStructure2011,
  title = {Simulation of Electronic Structure {{Hamiltonians}} Using Quantum Computers},
  author = {Whitfield, James D. and Biamonte, Jacob and {Aspuru-Guzik}, Al{\'a}n},
  year = 2011,
  month = mar,
  journal = {Molecular Physics},
  volume = {109},
  number = {5},
  pages = {735--750},
  issn = {0026-8976, 1362-3028},
  doi = {10.1080/00268976.2011.552441},
  urldate = {2020-09-15},
  langid = {english}
}

@article{klopperinitio1995,
  title = {An {\emph{Ab Initio}} Derived Torsional Potential Energy Surface for ({{H$_{2}$O}})$_{3}$. {{II}}. {{Benchmark}} Studies and Interaction Energies},
  author = {Klopper, Wim and Sch{\"u}tz, Martin and L{\"u}thi, Hans P. and Leutwyler, Samuel},
  year = 1995,
  month = jul,
  journal = {The Journal of Chemical Physics},
  volume = {103},
  number = {3},
  pages = {1085--1098},
  issn = {0021-9606, 1089-7690},
  doi = {10.1063/1.470701},
  urldate = {2025-10-23},
  langid = {english}
}

@article{hattigExplicitly2012,
  title = {Explicitly {{Correlated Electrons}} in {{Molecules}}},
  author = {H{\"a}ttig, Christof and Klopper, Wim and K{\"o}hn, Andreas and Tew, David P.},
  year = 2012,
  month = jan,
  journal = {Chemical Reviews},
  volume = {112},
  number = {1},
  pages = {4--74},
  issn = {0009-2665, 1520-6890},
  doi = {10.1021/cr200168z},
  urldate = {2025-10-23},
  langid = {english}
}

@article{shepherdConvergence2012,
  title = {Convergence of Many-Body Wave-Function Expansions Using a Plane-Wave Basis: {{From}} Homogeneous Electron Gas to Solid State Systems},
  shorttitle = {Convergence of Many-Body Wave-Function Expansions Using a Plane-Wave Basis},
  author = {Shepherd, James J. and Gr{\"u}neis, Andreas and Booth, George H. and Kresse, Georg and Alavi, Ali},
  year = 2012,
  month = jul,
  journal = {Physical Review B},
  volume = {86},
  number = {3},
  pages = {035111},
  issn = {1098-0121, 1550-235X},
  doi = {10.1103/PhysRevB.86.035111},
  urldate = {2025-10-23},
  copyright = {http://link.aps.org/licenses/aps-default-license},
  langid = {english}
}

@article{yanaiCanonical2006,
  title = {Canonical Transformation Theory for Multireference Problems},
  author = {Yanai, Takeshi and Chan, Garnet Kin-Lic},
  year = 2006,
  month = may,
  journal = {The Journal of Chemical Physics},
  volume = {124},
  number = {19},
  pages = {194106},
  issn = {0021-9606, 1089-7690},
  doi = {10.1063/1.2196410},
  urldate = {2025-01-09},
  langid = {english}
}

@article{hirschfelderRemoval1963,
  title = {Removal of {{Electron}}---{{Electron Poles}} from {{Many-Electron Hamiltonians}}},
  author = {Hirschfelder, Joseph O.},
  year = 1963,
  month = dec,
  journal = {The Journal of Chemical Physics},
  volume = {39},
  number = {11},
  pages = {3145--3146},
  issn = {0021-9606, 1089-7690},
  doi = {10.1063/1.1734157},
  urldate = {2025-10-23},
  langid = {english}
}

@article{kutzelnigg12Dependent1985,
  title = {$r_{12}$-{{Dependent}} Terms in the Wave Function as Closed Sums of Partial Wave Amplitudes for Large l},
  author = {Kutzelnigg, Werner},
  year = 1985,
  month = dec,
  journal = {Theoretica Chimica Acta},
  volume = {68},
  number = {6},
  pages = {445--469},
  issn = {0040-5744, 1432-2234},
  doi = {10.1007/BF00527669},
  urldate = {2025-10-23},
  copyright = {http://www.springer.com/tdm},
  langid = {english}
}

@article{magnussonEfficient2024,
  title = {Towards {{Efficient Quantum Computing}} for {{Quantum Chemistry}}: {{Reducing Circuit Complexity}} with {{Transcorrelated}} and {{Adaptive Ansatz Techniques}}},
  shorttitle = {Towards {{Efficient Quantum Computing}} for {{Quantum Chemistry}}},
  author = {Magnusson, Erika and Fitzpatrick, Aaron and Knecht, Stefan and Rahm, Martin and Dobrautz, Werner},
  year = 2024,
  journal = {Faraday Discussions},
  eprint = {2402.16659},
  primaryclass = {quant-ph},
  pages = {402-428},
  issn = {1359-6640, 1364-5498},
  doi = {10.1039/D4FD00039K},
  urldate = {2024-10-09},
  archiveprefix = {arXiv},
  langid = {english}
}

@article{filipDeterministic2025,
  title = {Deterministic {{Optimisation}} of {{Jastrow Factors}}},
  author = {Filip, Maria-Andreea and Christlmaier, Evelin M. C. and Haupt, J. Philip and Kats, Daniel and R{\'i}os, Pablo L{\'o}pez and Alavi, Ali},
  year = 2025,
  month = jun,
  pages = {arXiv:2506.04895},
  eprint = {2506.04895},
  primaryclass = {physics},
  journal = {arXiv},
  doi = {10.48550/arXiv.2506.04895},
  urldate = {2026-02-09},
  archiveprefix = {arXiv},
  langid = {english}
}

@article{mottaQuantum2020,
  title = {Quantum Simulation of Electronic Structure with a Transcorrelated {{Hamiltonian}}: Improved Accuracy with a Smaller Footprint on the Quantum Computer},
  shorttitle = {Quantum Simulation of Electronic Structure with a Transcorrelated {{Hamiltonian}}},
  author = {Motta, Mario and Gujarati, Tanvi P. and Rice, Julia E. and Kumar, Ashutosh and Masteran, Conner and Latone, Joseph A. and Lee, Eunseok and Valeev, Edward F. and Takeshita, Tyler Y.},
  year = 2020,
  journal = {Physical Chemistry Chemical Physics},
  volume = {22},
  number = {42},
  pages = {24270--24281},
  issn = {1463-9076, 1463-9084},
  doi = {10.1039/D0CP04106H},
  urldate = {2024-11-08},
  langid = {english}
}

@article{leeEvenMoreEfficient2021,
  title = {Even {{More Efficient Quantum Computations}} of {{Chemistry Through Tensor Hypercontraction}}},
  author = {Lee, Joonho and Berry, Dominic W. and Gidney, Craig and Huggins, William J. and McClean, Jarrod R. and Wiebe, Nathan and Babbush, Ryan},
  year = 2021,
  month = jul,
  journal = {PRX Quantum},
  volume = {2},
  number = {3},
  pages = {030305},
  issn = {2691-3399},
  doi = {10.1103/PRXQuantum.2.030305},
  urldate = {2024-06-24},
  langid = {english}
}

@book{bhatiaMatrix1997,
  title = {Matrix Analysis},
  author = {Bhatia, Rajendra},
  year = 1997,
  series = {Graduate Texts in Mathematics},
  number = {169},
  publisher = {Springer},
  address = {New York Berlin Paris [etc.]},
  isbn = {978-0-387-94846-1},
  langid = {english},
  lccn = {512.943 4}
}

@article{jonesLowoverheadConstructionsFaulttolerant2013,
  title = {Low-Overhead Constructions for the Fault-Tolerant {{Toffoli}} Gate},
  author = {Jones, Cody},
  year = 2013,
  month = feb,
  journal = {Physical Review A},
  volume = {87},
  number = {2},
  pages = {022328},
  issn = {1050-2947, 1094-1622},
  doi = {10.1103/PhysRevA.87.022328},
  urldate = {2020-07-09},
  langid = {english}
}

@article{berryQubitization2019,
  title = {Qubitization of {{Arbitrary Basis Quantum Chemistry Leveraging Sparsity}} and {{Low Rank Factorization}}},
  author = {Berry, Dominic W. and Gidney, Craig and Motta, Mario and McClean, Jarrod R. and Babbush, Ryan},
  year = 2019,
  month = dec,
  journal = {Quantum},
  volume = {3},
  eprint = {1902.02134},
  primaryclass = {physics, physics:quant-ph},
  pages = {208},
  issn = {2521-327X},
  doi = {10.22331/q-2019-12-02-208},
  urldate = {2024-07-08},
  archiveprefix = {arXiv}
}

@misc{DarpaQB,
  author = {},
  title = {{QB-GSEE-Benchmark}},
  year = {2025},
  publisher = {GitHub},
  howpublished = {\url{https://github.com/isi-usc-edu/qb-gsee-benchmark/}},
  note = {Accessed: 2025-10-21}
}

@article{gruneisPerspective2017,
  title = {Perspective: {{Explicitly}} Correlated Electronic Structure Theory for Complex Systems},
  shorttitle = {Perspective},
  author = {Gr{\"u}neis, Andreas and Hirata, So and Ohnishi, Yu-ya and {Ten-no}, Seiichiro},
  year = 2017,
  month = feb,
  journal = {The Journal of Chemical Physics},
  volume = {146},
  number = {8},
  pages = {080901},
  issn = {0021-9606, 1089-7690},
  doi = {10.1063/1.4976974},
  urldate = {2025-09-02},
  langid = {english}
}

@article{hylleraasNeue1929,
  title = {{Neue Berechnung der Energie des Heliums im Grundzustande, sowie des tiefsten Terms von Ortho-Helium}},
  author = {Hylleraas, Egil A.},
  year = 1929,
  month = may,
  journal = {Zeitschrift f\"ur Physik},
  volume = {54},
  number = {5-6},
  pages = {347--366},
  issn = {1434-6001, 1434-601X},
  doi = {10.1007/BF01375457},
  urldate = {2025-10-21},
  copyright = {http://www.springer.com/tdm},
  langid = {ngerman}
}

@book{nielsenQuantumComputationQuantum2010,
  title = {Quantum Computation and Quantum Information},
  author = {Nielsen, Michael A and Chuang, Isaac L},
  year = 2010,
  publisher = {Cambridge University Press},
  address = {Cambridge; New York},
  urldate = {2020-03-20},
  isbn = {978-1-107-00217-3 978-0-511-97666-7 978-1-282-96729-8},
  langid = {english}
}

@article{boysdetermination1969,
  title = {The Determination of Energies and Wavefunctions with Full Electronic Correlation},
  author = {Boys, Samuel Francis and Handy, Nicholas Charles},
  year = 1969,
  month = apr,
  journal = {Proceedings of the Royal Society of London. A. Mathematical and Physical Sciences},
  volume = {310},
  number = {1500},
  pages = {43--61},
  issn = {0080-4630, 2053-9169},
  doi = {10.1098/rspa.1969.0061},
  urldate = {2026-03-30},
  copyright = {https://royalsociety.org/journals/ethics-policies/data-sharing-mining/},
  langid = {english}
}

@article{liCLASS2014,
  title = {A class of efficient quantum incrementer gates for quantum circuit synthesis},
  author = {Li, Xiaoyu and Yang, Guowu and Torres, Carlos Manuel and Zheng, Desheng and Wang, Kang L.},
  year = {2014},
  month = jan,
  journal = {International Journal of Modern Physics B},
  volume = {28},
  number = {01},
  pages = {1350191},
  issn = {0217-9792, 1793-6578},
  doi = {10.1142/S0217979213501919},
  urldate = {2025-09-05},
  langid = {english}
}

@article{costaOptimal2022,
  title = {Optimal {{Scaling Quantum Linear-Systems Solver}} via {{Discrete Adiabatic Theorem}}},
  author = {Costa, Pedro C.S. and An, Dong and Sanders, Yuval R. and Su, Yuan and Babbush, Ryan and Berry, Dominic W.},
  year = {2022},
  month = oct,
  journal = {PRX Quantum},
  volume = {3},
  number = {4},
  publisher = {American Physical Society (APS)},
  issn = {2691-3399},
  doi = {10.1103/prxquantum.3.040303},
  urldate = {2025-07-17},
  copyright = {https://creativecommons.org/licenses/by/4.0/},
  langid = {english}
}

@article{chakravortyGroundstate1993,
  title = {Ground-State Correlation Energies for Atomic Ions with 3 to 18 Electrons},
  author = {Chakravorty, Subhas J. and Gwaltney, Steven R. and Davidson, Ernest R. and Parpia, Farid A. and P Fischer, Charlotte Froese},
  year = {1993},
  month = may,
  journal = {Physical Review A},
  volume = {47},
  number = {5},
  pages = {3649--3670},
  issn = {1050-2947, 1094-1622},
  doi = {10.1103/PhysRevA.47.3649},
  urldate = {2025-06-06},
  copyright = {http://link.aps.org/licenses/aps-default-license},
  langid = {english}
}

@article{dobrautzInitio2024,
  title = {Ab {{Initio Transcorrelated Method}} Enabling Accurate {{Quantum Chemistry}} on Near-Term {{Quantum Hardware}}},
  author = {Dobrautz, Werner and Sokolov, Igor O. and Liao, Ke and R{\'i}os, Pablo L{\'o}pez and Rahm, Martin and Alavi, Ali and Tavernelli, Ivano},
  journal={arXiv},
  year={2023},
  pages = {arXiv:2303.02007},
  number = {2303.02007},
  eprint = {2303.02007},
  publisher = {arXiv},
  urldate = {2024-09-20},
  archiveprefix = {arXiv},
  langid = {english}
}

@article{needsVariational2020,
  title = {Variational and Diffusion Quantum {{Monte Carlo}} Calculations with the {{CASINO}} Code},
  author = {Needs, R. J. and Towler, M. D. and Drummond, N. D. and L{\'o}pez R{\'i}os, P. and Trail, J. R.},
  year = {2020},
  month = apr,
  journal = {The Journal of Chemical Physics},
  volume = {152},
  number = {15},
  pages = {154106},
  issn = {0021-9606, 1089-7690},
  doi = {10.1063/1.5144288},
  urldate = {2025-06-09},
  langid = {english}
}

@article{drummondJastrow2004,
  title = {Jastrow Correlation Factor for Atoms, Molecules, and Solids},
  author = {Drummond, N. D. and Towler, M. D. and Needs, R. J.},
  year = {2004},
  month = dec,
  journal = {Physical Review B},
  volume = {70},
  number = {23},
  pages = {235119},
  issn = {1098-0121, 1550-235X},
  doi = {10.1103/PhysRevB.70.235119},
  urldate = {2025-03-18},
  copyright = {http://link.aps.org/licenses/aps-default-license},
  langid = {english}
}

@inproceedings{lowQuantum2024,
  title = {Quantum {{Eigenvalue Processing}}},
  booktitle = {2024 {{IEEE}} 65th {{Annual Symposium}} on {{Foundations}} of {{Computer Science}} ({{FOCS}})},
  author = {Low, Guang Hao and Su, Yuan},
  year = {2024},
  month = oct,
  pages = {1051--1062},
  publisher = {IEEE},
  address = {Chicago, IL, USA},
  doi = {10.1109/FOCS61266.2024.00070},
  urldate = {2025-02-26},
  copyright = {https://doi.org/10.15223/policy-029},
  isbn = {979-8-3315-1674-1},
  langid = {english}
}

@article{babbushEncoding2018,
  title = {Encoding {{Electronic Spectra}} in {{Quantum Circuits}} with {{Linear T Complexity}}},
  author = {Babbush, Ryan and Gidney, Craig and Berry, Dominic W. and Wiebe, Nathan and McClean, Jarrod and Paler, Alexandru and Fowler, Austin and Neven, Hartmut},
  year = {2018},
  month = oct,
  journal = {Physical Review X},
  volume = {8},
  number = {4},
  pages = {041015},
  issn = {2160-3308},
  doi = {10.1103/PhysRevX.8.041015},
  urldate = {2024-05-16},
  langid = {english}
}

@article{sokolovOrders2023,
  title = {Orders of Magnitude Increased Accuracy for Quantum Many-Body Problems on Quantum Computers via an Exact Transcorrelated Method},
  author = {Sokolov, Igor O. and Dobrautz, Werner and Luo, Hongjun and Alavi, Ali and Tavernelli, Ivano},
  year = 2023,
  month = jun,
  journal = {Physical Review Research},
  volume = {5},
  number = {2},
  pages = {023174},
  issn = {2643-1564},
  doi = {10.1103/PhysRevResearch.5.023174},
  urldate = {2026-03-25},
  langid = {english}
}

@article{shaoComputing2020,
  title = {Computing Eigenvalues of Diagonalizable Matrices in a Quantum Computer},
  author = {Shao, Changpeng},
  year = 2020,
  month = sep,
  pages = {arXiv:1912.08015},
  eprint = {1912.08015},
  primaryclass = {quant-ph},
  publisher = {arXiv},
  doi = {10.48550/arXiv.1912.08015},
  urldate = {2026-03-25},
  archiveprefix = {arXiv},
  journal={arXiv}
}

@article{fournaisSharp2005,
  title = {Sharp {{Regularity Results}} for {{Coulombic Many-Electron Wave Functions}}},
  author = {Fournais, S{\o}ren and {Hoffmann-Ostenhof}, Maria and {Hoffmann-Ostenhof}, Thomas and S{\o}rensen, Thomas {\O}stergaard},
  year = 2005,
  month = apr,
  journal = {Communications in Mathematical Physics},
  volume = {255},
  number = {1},
  pages = {183--227},
  issn = {0010-3616, 1432-0916},
  doi = {10.1007/s00220-004-1257-6},
  urldate = {2024-09-27},
  copyright = {http://www.springer.com/tdm},
  langid = {english}
}

@article{ten-nofeasible2000,
  title = {A Feasible Transcorrelated Method for Treating Electronic Cusps Using a Frozen {{Gaussian}} Geminal},
  author = {{Ten-no}, Seiichiro},
  year = 2000,
  month = nov,
  journal = {Chemical Physics Letters},
  volume = {330},
  number = {1-2},
  pages = {169--174},
  issn = {00092614},
  doi = {10.1016/S0009-2614(00)01066-6},
  urldate = {2025-11-05},
  langid = {english}
}

@article{christlmaierxTC2023,
  title = {{{xTC}}: {{An}} Efficient Treatment of Three-Body Interactions in Transcorrelated Methods},
  shorttitle = {{{xTC}}},
  author = {Christlmaier, Evelin Martine Corvid and Schraivogel, Thomas and L{\'o}pez R{\'i}os, Pablo and Alavi, Ali and Kats, Daniel},
  year = 2023,
  month = jul,
  journal = {The Journal of Chemical Physics},
  volume = {159},
  number = {1},
  pages = {014113},
  issn = {0021-9606, 1089-7690},
  doi = {10.1063/5.0154445},
  urldate = {2025-12-10},
  langid = {english}
}

@article{schraivogelTranscorrelated2021,
  title = {Transcorrelated Coupled Cluster Methods},
  author = {Schraivogel, Thomas and Cohen, Aron J. and Alavi, Ali and Kats, Daniel},
  year = 2021,
  month = nov,
  journal = {The Journal of Chemical Physics},
  volume = {155},
  number = {19},
  pages = {191101},
  issn = {0021-9606, 1089-7690},
  doi = {10.1063/5.0072495},
  urldate = {2026-03-20},
  langid = {english}
}

@article{schraivogelTranscorrelated2023,
  title = {Transcorrelated Coupled Cluster Methods. {{II}}. {{Molecular}} Systems},
  author = {Schraivogel, Thomas and Christlmaier, Evelin Martine Corvid and L{\'o}pez R{\'i}os, Pablo and Alavi, Ali and Kats, Daniel},
  year = 2023,
  month = jun,
  journal = {The Journal of Chemical Physics},
  volume = {158},
  number = {21},
  pages = {214106},
  issn = {0021-9606, 1089-7690},
  doi = {10.1063/5.0151412},
  urldate = {2026-03-20},
  langid = {english}
}

@article{torheydenUniversal2009,
  title = {Universal Perturbative Explicitly Correlated Basis Set Incompleteness Correction},
  author = {Torheyden, Martin and Valeev, Edward F.},
  year = 2009,
  month = nov,
  journal = {The Journal of Chemical Physics},
  volume = {131},
  number = {17},
  pages = {171103},
  issn = {0021-9606, 1089-7690},
  doi = {10.1063/1.3254836},
  urldate = {2026-03-18},
  langid = {english}
}

@article{mcardleImproving2020,
  title = {Improving the Accuracy of Quantum Computational Chemistry Using the Transcorrelated Method},
  author = {McArdle, Sam and Tew, David P.},
  year = 2020,
  month = jun,
  pages = {arXiv:2006.11181},
  eprint = {2006.11181},
  primaryclass = {quant-ph},
  publisher = {arXiv},
  doi = {10.48550/arXiv.2006.11181},
  urldate = {2026-02-19},
  archiveprefix = {arXiv},
  langid = {english},
  journal={arXiv}
}

@article{schleichImproving2022,
  title = {Improving the Accuracy of the Variational Quantum Eigensolver for Molecular Systems by the Explicitly-Correlated Perturbative [2]{{{\textsubscript{R12}}}} {\textbf{-}} Correction},
  author = {Schleich, Philipp and Kottmann, Jakob S. and {Aspuru-Guzik}, Al{\'a}n},
  year = 2022,
  journal = {Physical Chemistry Chemical Physics},
  volume = {24},
  number = {22},
  pages = {13550--13564},
  issn = {1463-9076, 1463-9084},
  doi = {10.1039/D2CP00247G},
  urldate = {2026-03-16},
  langid = {english}
}

@article{tillyVariationalQuantumEigensolver2022,
  title = {The {{Variational Quantum Eigensolver}}: {{A}} Review of Methods and Best Practices},
  shorttitle = {The {{Variational Quantum Eigensolver}}},
  author = {Tilly, Jules and Chen, Hongxiang and Cao, Shuxiang and Picozzi, Dario and Setia, Kanav and Li, Ying and Grant, Edward and Wossnig, Leonard and Rungger, Ivan and Booth, George H. and Tennyson, Jonathan},
  year = 2022,
  month = nov,
  journal = {Physics Reports},
  volume = {986},
  pages = {1--128},
  issn = {03701573},
  doi = {10.1016/j.physrep.2022.08.003},
  urldate = {2022-12-19},
  langid = {english}
}

@article{arponenmethod1982,
  title = {The Method of Stationary Cluster Amplitudes and the Phase Transition in the {{Lipkin}} Pseudospin Model},
  author = {Arponen, J},
  year = 1982,
  month = aug,
  journal = {Journal of Physics G: Nuclear Physics},
  volume = {8},
  number = {8},
  pages = {L129-L134},
  issn = {0305-4616},
  doi = {10.1088/0305-4616/8/8/004},
  urldate = {2026-03-30}
}

@article{arponenVariational1983,
  title = {Variational Principles and Linked-Cluster Exp {{S}} Expansions for Static and Dynamic Many-Body Problems},
  author = {Arponen, Jouko},
  year = 1983,
  month = dec,
  journal = {Annals of Physics},
  volume = {151},
  number = {2},
  pages = {311--382},
  issn = {00034916},
  doi = {10.1016/0003-4916(83)90284-1},
  urldate = {2026-03-30},
  copyright = {https://www.elsevier.com/tdm/userlicense/1.0/},
  langid = {english}
}

@article{lowdinstability1983,
  title = {On the Stability Problem of a Pair of Adjoint Operators},
  author = {L{\"o}wdin, Per-Olov},
  year = 1983,
  month = jan,
  journal = {Journal of Mathematical Physics},
  volume = {24},
  number = {1},
  pages = {70--87},
  issn = {0022-2488, 1089-7658},
  doi = {10.1063/1.525604},
  urldate = {2026-03-30},
  langid = {english}
}

@article{kongExplicitly2012,
  title = {Explicitly {{Correlated R12}}/{{F12 Methods}} for {{Electronic Structure}}},
  author = {Kong, Liguo and Bischoff, Florian A. and Valeev, Edward F.},
  year = 2012,
  month = jan,
  journal = {Chemical Reviews},
  volume = {112},
  number = {1},
  pages = {75--107},
  issn = {0009-2665, 1520-6890},
  doi = {10.1021/cr200204r},
  urldate = {2024-09-24},
  langid = {english}
}

@article{tsuneyukiTranscorrelated2008,
  title = {Transcorrelated {{Method}}: {{Another Possible Way}} towards {{Electronic Structure Calculation}} of {{Solids}}},
  shorttitle = {Transcorrelated {{Method}}},
  author = {Tsuneyuki, Shinji},
  year = 2008,
  journal = {Progress of Theoretical Physics Supplement},
  volume = {176},
  pages = {134--142},
  issn = {0375-9687},
  doi = {10.1143/PTPS.176.134},
  urldate = {2026-03-30},
  langid = {english}
}

@article{jastrowManyBody1955,
  title = {Many-{{Body Problem}} with {{Strong Forces}}},
  author = {Jastrow, Robert},
  year = 1955,
  month = jun,
  journal = {Physical Review},
  volume = {98},
  number = {5},
  pages = {1479--1484},
  issn = {0031-899X},
  doi = {10.1103/PhysRev.98.1479},
  urldate = {2024-12-18},
  copyright = {http://link.aps.org/licenses/aps-default-license},
  langid = {english}
}

@article{kutzelniggNormal1997,
  title = {Normal Order and Extended {{Wick}} Theorem for a Multiconfiguration Reference Wave Function},
  author = {Kutzelnigg, Werner and Mukherjee, Debashis},
  year = 1997,
  month = jul,
  journal = {The Journal of Chemical Physics},
  volume = {107},
  number = {2},
  pages = {432--449},
  issn = {0021-9606, 1089-7690},
  doi = {10.1063/1.474405},
  urldate = {2025-11-13},
  langid = {english}
}

@article{sunRecent2020,
  title = {Recent Developments in the {{P}}{\textsc{y}}{{SCF}} Program Package},
  shorttitle = {Recent Developments in the {{P}}},
  author = {Sun, Qiming and Zhang, Xing and Banerjee, Samragni and Bao, Peng and Barbry, Marc and Blunt, Nick S. and Bogdanov, Nikolay A. and Booth, George H. and Chen, Jia and Cui, Zhi-Hao and Eriksen, Janus J. and Gao, Yang and Guo, Sheng and Hermann, Jan and Hermes, Matthew R. and Koh, Kevin and Koval, Peter and Lehtola, Susi and Li, Zhendong and Liu, Junzi and Mardirossian, Narbe and McClain, James D. and Motta, Mario and Mussard, Bastien and Pham, Hung Q. and Pulkin, Artem and Purwanto, Wirawan and Robinson, Paul J. and Ronca, Enrico and Sayfutyarova, Elvira R. and Scheurer, Maximilian and Schurkus, Henry F. and Smith, James E. T. and Sun, Chong and Sun, Shi-Ning and Upadhyay, Shiv and Wagner, Lucas K. and Wang, Xiao and White, Alec and Whitfield, James Daniel and Williamson, Mark J. and Wouters, Sebastian and Yang, Jun and Yu, Jason M. and Zhu, Tianyu and Berkelbach, Timothy C. and Sharma, Sandeep and Sokolov, Alexander Yu. and Chan, Garnet Kin-Lic},
  year = 2020,
  month = jul,
  journal = {The Journal of Chemical Physics},
  volume = {153},
  number = {2},
  pages = {024109},
  issn = {0021-9606, 1089-7690},
  doi = {10.1063/5.0006074},
  urldate = {2026-02-27},
  langid = {english}
}

@article{dobrautzCompact2019,
  title = {Compact Numerical Solutions to the Two-Dimensional Repulsive {{Hubbard}} Model Obtained via Nonunitary Similarity Transformations},
  author = {Dobrautz, Werner and Luo, Hongjun and Alavi, Ali},
  year = 2019,
  month = feb,
  journal = {Physical Review B},
  volume = {99},
  number = {7},
  pages = {075119},
  issn = {2469-9950, 2469-9969},
  doi = {10.1103/PhysRevB.99.075119},
  urldate = {2026-02-19},
  langid = {english}
}

@article{liaoefficient2021,
  title = {Towards Efficient and Accurate {\emph{Ab Initio}} Solutions to Periodic Systems via Transcorrelation and Coupled Cluster Theory},
  author = {Liao, Ke and Schraivogel, Thomas and Luo, Hongjun and Kats, Daniel and Alavi, Ali},
  year = 2021,
  month = jul,
  journal = {Physical Review Research},
  volume = {3},
  number = {3},
  pages = {033072},
  issn = {2643-1564},
  doi = {10.1103/PhysRevResearch.3.033072},
  urldate = {2026-04-08},
  langid = {english}
}

@article{liaoDensity2023,
  title = {Density {{Matrix Renormalization Group}} for {{Transcorrelated Hamiltonians}}: {{Ground}} and {{Excited States}} in {{Molecules}}},
  shorttitle = {Density {{Matrix Renormalization Group}} for {{Transcorrelated Hamiltonians}}},
  author = {Liao, Ke and Zhai, Huanchen and Christlmaier, Evelin Martine Corvid and Schraivogel, Thomas and R{\'i}os, Pablo L{\'o}pez and Kats, Daniel and Alavi, Ali},
  year = 2023,
  month = mar,
  journal = {Journal of Chemical Theory and Computation},
  volume = {19},
  number = {6},
  pages = {1734--1743},
  issn = {1549-9618, 1549-9626},
  doi = {10.1021/acs.jctc.2c01207},
  urldate = {2026-04-08},
  copyright = {https://doi.org/10.15223/policy-029},
  langid = {english}
}

@PREAMBLE{
 "\providecommand{\noopsort}[1]{}" 
 # "\providecommand{\singleletter}[1]{#1}%" 
}

\end{document}